\documentclass[manuscript, screen]{acmart}

\setcopyright{none}

\usepackage[dvipsnames]{xcolor}
\usepackage{enumitem}
\usepackage{wrapfig}
\usepackage{hhline}
\usepackage{listings}

\newcommand{\acol}[1]{\ensuremath{a_{#1}}}
\newcommand{\abcol}[1]{\ensuremath{A_{#1}}}
\newcommand{\vcol}[1]{\ensuremath{v_{#1}}}

\usepackage{mathtools}
\DeclarePairedDelimiter{\norm}{\lVert}{\rVert}
\DeclarePairedDelimiter{\abs}{\lvert}{\rvert}
\newcommand{\fnorm}[1]{\ensuremath{\norm{#1}_F}}

\DeclareMathOperator{\diag}{diag}
\DeclareMathOperator{\off}{off}
\DeclareMathOperator{\sign}{sign}

\newcommand{\C}{\ensuremath{\mathbb{C}}}
\newcommand{\Cmn}{\ensuremath{\mathbb{C}^{m \times n}}}

\newcommand{\Cnn}{\ensuremath{\mathbb{C}^{n \times n}}}

\newcommand{\nb}{\ensuremath{nb}}
\newcommand{\bc}{\ensuremath{\ell}}
\newcommand{\mb}{\ensuremath{mb}}

\newcommand{\wh}[1]{\widehat{#1}}

\newcommand{\Cmnb}{\ensuremath{\mathbb{C}^{m \times \nb}}}
\newcommand{\Cnbnb}{\ensuremath{\mathbb{C}^{\nb \times \nb}}}
\newcommand{\Ctnbtnb}{\ensuremath{\mathbb{C}^{2\nb \times 2\nb}}}

\renewcommand{\Sigma}{\varSigma}

\usepackage[ruled,vlined,linesnumbered]{algorithm2e}
\DontPrintSemicolon
\SetKwInput{KwParam}{Parameters}
\newcommand{\false}{\texttt{False}}
\newcommand{\true}{\texttt{True}}
\newcommand{\mnot}{\textbf{not}}
\newcommand{\mand}{\textbf{and}}

\renewenvironment{quote}
  {\list{}{\rightmargin=.3cm \leftmargin=.3cm}%
   \item\relax}
  {\endlist}

\AtBeginDocument{%
  }

\renewcommand{\tt}[1]{\texttt{#1}}

\newcommand{\bld}[1]{\textbf{#1}}

\newcommand{\condt}[1]{\kappa_2({#1})}

\newcommand{\pluseq}{\mathrel{+}=}

\usepackage{calc}
\usepackage[capitalise]{cleveref}
\begin{document}

\title[Batch SVD on GPUs]{An Efficient Batch Solver for the Singular Value Decomposition on GPUs}

\author{Ahmad Abdelfattah}
\email{ahmad@icl.utk.edu}
\orcid{0000-0001-5054-4784}
\affiliation{%
  \institution{University of Tennessee}
  \city{Knoxville}
  \state{TN}
  \country{USA}
}

\author{Massimiliano Fasi}
\email{m.fasi@leeds.ac.uk}
\orcid{0000-0002-6015-391X}
\affiliation{%
  \institution{University of Leeds}
  \city{Leeds}
  \country{UK}}
\email{m.fasi@leeds.ac.uk}

\renewcommand{\shortauthors}{Abdelfattah and Fasi}

\begin{abstract}

The singular value decomposition (SVD) is a powerful tool in modern numerical linear algebra, which underpins computational methods such as principal component analysis (PCA), low-rank approximations, and randomized algorithms. Many practical scenarios require solving numerous small SVD problems, a regime generally referred to as \emph{batch SVD}. Existing programming models can handle this efficiently on parallel CPU architectures, but high-performance solutions for GPUs remain immature. A GPU-oriented batch SVD solver is introduced. This solver exploits the one-sided Jacobi algorithm to exploit fine-grained parallelism, and a number of algorithmic and design optimizations achieve unmatched performance. Starting from a baseline solver, a sequence of optimizations is applied to obtain incremental performance gains. Numerical experiments show that the new solver is robust across problems with different numerical properties, matrix shapes, and arithmetic precisions. Performance benchmarks on both NVIDIA and AMD systems show significant performance speedups over vendor solutions as well as existing open-source solvers. 
\end{abstract}



\maketitle

\section{Computing the Singular Value Decomposition}
\label{sec:intro}
The singular value decomposition (SVD) is one of the most powerful and versatile tools in modern linear algebra.
Its roots trace back to the work of classical mathematicians~\cite{stew93}, from James Sylvester (1814--1897), Eugenio Beltrami (1835--1899), and Camille Jordan (1838--1921) to Erhard Schmidt (1876-1959) and Hermann Weyl (1885--1955).
In essence, the SVD constructs orthonormal bases for the row and column spaces of a matrix.
These bases reveal rank information and can be used to construct optimal low-rank approximations, which are necessary in a wide range of applications.

In numerical linear algebra, the singular values are used to estimate the effective rank~\cite{klla80} and the 2-norm condition number of a matrix $A$.
In numerical algorithms, the SVD can be used to compute the Moore--Penrose pseudoinverse of a matrix~\cite{moor20,penr55}, which can be used to solve linear least squares problems~\cite{penr56} and has been applied to the processing of images~\cite{anpa76} as well as genomic data~\cite{algo04}.
In data science and statistics, the geometric interpretation of the SVD underlies principal component analysis~\cite{pear01,hote33,joca16,moor81} and other techniques that rely on understanding variance and structure in data~\cite{cugh15}.
Finally, Eckart and Young~\cite{ecyo36} have shown that truncating the SVD
provides the best low-rank approximation to $A$.
Such low-rank approximations are essential in dimensionality reduction and data compression techniques, and they play an important role in reducing the complexity of neural networks~\cite{dzbl14} and large language models~\cite{wzwz25}.
A selection of other interesting applications can be found in the short survey by Martin and Porter~\cite{mapo12}.


Here, we focus on the \emph{reduced} or \emph{economy-size} SVD, which we now define.
Every matrix $A \in \Cmn$ can be written as
\begin{equation}\label{eq:svd}
    A = U \Sigma V^{H},
\end{equation}
where $U \in \C^{m \times \min(m,n)}$ and $V \in \C^{n \times \min(m,n)}$ have orthonormal columns, $\Sigma  \in \C^{\min(m,n) \times \min(m,n)}$ is diagonal with non-negative real entries, and $V^H$ denotes the conjugate transpose of $V$.
The columns of $U$ and $V$ are the \emph{left singular vectors} and \emph{right singular vectors} of $A$, respectively, and the values along the diagonal of $\Sigma$ are its \emph{singular values}.
The singular values often appear along the diagonal of $\Sigma$ in descending order, so that
\begin{equation*}
    \sigma_1 \geq \sigma_{2} \geq \ldots \geq \sigma_{\min(m,n)} \geq 0.
\end{equation*}

The SVD is tightly connected to the Hermitian eigenvalue problem, because
\begin{equation}\label{eq:gram-matrices}
  A^HA = V \Sigma U^H U \Sigma V^H = V \Sigma^2 V^H,\qquad
  AA^H = U \Sigma V^H V \Sigma U^H = U \Sigma^2 U^H.
\end{equation}
In other words, the right and left singular vectors of $A$ are the eigenvectors of the Gram matrices $A^HA$ and $AA^H$, respectively.
The two Gram matrices have the same eigenvalues, which are the squares of the singular values of $A$.
This property is the algorithmic foundation of the efficient design introduced in this paper.

In view of~\eqref{eq:gram-matrices}, computing the SVD can be reduced to the solution of a Hermitian eigenvalue problem: if we solve $A^HA= V \Sigma^2 V^H$, then the relation $AV = U\Sigma$ shows that the singular values of $A$ are the norms of the columns of $AV$, and thus $U$ can be recovered from the identity $U = AV\Sigma^{-1}$.
Equivalently, we can solve the Hermitian eigenvalue problem $AA^H = U \Sigma^2 U^H$ and use the definition of the SVD to obtain $V$.
In the literature, we can identify two main families of efficient and practical algorithms to compute the SVD.

On the one hand, we have algorithms that reduce $A$ to a simpler form.
They usually perform three phases.
\begin{enumerate}[nolistsep,label=(\roman*),ref=(\roman*)]
\item \bld{First}, they reduce the input matrix $A$ to a \emph{bi-diagonal} form $A = U_0BV_0^H$. Computationally, this bi-diagonalization phase is usually the most expensive, and much effort has been dedicated to accelerating it on parallel architectures, including one-stage~\cite{lana09} and two-stage~\cite{lang96,grla99} algorithms.\label{it:phase1}

\item \bld{Second}, they use an SVD solver to compute $U_1$ and $V_1$ such that $B = U_1\Sigma V_1^H$. Unlike~\ref{it:phase1}, this second phase is an iterative process in general. Customary techniques for this diagonalization are the QR algorithm, the Divide and Conquer (D\&C) algorithm, and the Multiple Relatively Robust Representations (MRRR) method.
These iterative solvers are relatively lightweight, compared with the bi-diagonalization scheme.
Most of them, however, are nontrivial to parallelize efficiently, especially on GPUs.\label{it:phase2}

\item \bld{Third}, they compute the singular vectors of $A$ as $U=U_0 U_1$ and $V = V_0 V_1$.
This final phase is only performed if the singular vectors are needed.
Computationally, this only requires matrix multiplications, which are highly optimized on all parallel architectures.\label{it:phase3}
\end{enumerate}
In view of~\ref{it:phase1}, all techniques that follow this idea are collectively known as \emph{bi-diagonalization algorithms}.

The alternative is to use \emph{Jacobi-type algorithms}, which are based on the Jacobi solver for Hermitian eigenproblems.
In contrast to bi-diagonalization algorithms, Jacobi methods operate directly on $A$ and do not require the preprocessing step to reduce it to bi-diagonal form.
There are two standard routes to compute the SVD using Jacobi-type algorithms:
\begin{itemize}[nolistsep]
    \item the two-sided approach, which applies plane rotations on both sides of $A$ until it is diagonalized; and
    \item the one-sided approach, which implicitly diagonalizes the Hermitian matrix $A^HA$ without explicitly forming it.
\end{itemize}
Jacobi algorithms are often slower than those based on bi-diagonalization, because they require substantially more floating point operations (FLOPs) than the alternatives mentioned above~\cite[sect.~5.3]{demm97}.
However, there has been a resurgence of interest in these methods because of their simplicity, ease of parallelization, and potential for higher accuracy on certain classes of problems~\cite{dghk18}.

In this paper, we consider the one-sided Jacobi algorithm for computing the SVD of many relatively small problems (i.e., a batch SVD) using GPUs. The developed batch SVD solver brings a number of novel algorithmic and design techniques that combine performance with numerical robustness, and is publicly available through the open-source MAGMA library~\cite{agullo2009numerical,abcg24}.
Here is a list of our contributions.
\begin{enumerate}[nolistsep]
    \item We present a GPU-accelerated batch SVD solver that outperforms the state of the art. The solver supports matrices of any shape and size, as long as the batch fits in the GPU memory. Unlike other open-source solutions, the solver supports both real and complex problems in 32-bit and 64-bit floating-point arithmetic. 
    \item We propose an algorithmic change in the original one-sided Jacobi algorithm that dramatically improves its performance on the GPU without sacrificing accuracy.
    \item We develop optimized GPU kernels for almost every computational stage in the one-sided Jacobi algorithm. The developed kernels enable a significant performance gain over the baseline implementation.
    \item We propose \emph{masked batch operations}, a mechanism for masking off computational stages of certain problems in the batch if they are detected to have converged. 
    \item We present comprehensive benchmarks for performance and numerical accuracy across different GPUs, different compute precisions, and different problem sizes.
\end{enumerate}

\section{Related Work}
\label{sec:related}

Dense matrix algorithms have been studied extensively for acceleration on GPUs, including the SVD and eigenvalue decomposition algorithms. We distinguish between batch algorithms, where many independent and relatively small matrices are considered, and non-batch algorithms, where a single relatively large matrix provides enough parallel work for the hardware.

Single-matrix workloads have been studied extensively in the literature.
Gates et al.~\cite{gates2018accelerating} developed a GPU SVD solver based on the two-stage reduction to a bi-diagonal form, followed by a parallel Divide and Conquer algorithm. Kabir et al.~\cite{kabir2017framework} developed an out-of-core SVD solver, where the matrix is so large that disk storage is required. Novaković~\cite{nova15} implemented a multi-level blocking Jacobi SVD algorithm for single and multi-GPU systems.
Some internal components can be considered batch kernels, but they are designed for a fixed block size that only serves the purpose of one large SVD, and so they do not provide a generic batch SVD functionality of an arbitrary shape.
Novaković and Singer~\cite{nosi20} took a similar approach for the generalized SVD problem.
Several block-Jacobi variants for Hermitian and indefinite problems have been proposed, including both block-oriented~\cite{hss10} and full-block~\cite{hss14} formulations aimed at improving convergence and exploiting BLAS 3 structure.

In scientific applications, one often needs to compute the SVD of many small matrices rather than of a single, large one. This situation arises, for example, in hierarchical matrix compression~\cite{btlk18}, image mosaic assemble~\cite{bpf15}, computation of separable filters for convolutional neural networks~\cite{kale15}, simulation of quantum lattice systems~\cite{hylz22}, and Tucker decomposition of tensors~\cite{lhkc24}.
If the matrices are small, they cannot saturate the resources available on the GPU, and computing their SVDs will under-utilize the hardware. In addition, it has been shown that using concurrent execution using parallel GPU streams/queues does not achieve an acceptable performance~\cite{abdelfattah2016performance}.
To leverage parallelism and maximize throughput, it is convenient to use \emph{batch} algorithms, which will compute multiple independent SVDs simultaneously.
Both bi-diagonalization and Jacobi-type algorithms have been adapted to support batched execution, with recent work showing significant speedups and improved scalability.

Batch linear algebra algorithms have been studied and developed for many matrix operations, including BLAS operations~\cite{abdelfattah2016performance,abdelfattah2017novel,abdelfattah2019fast} and the Cholesky~\cite{dhtd14,abdelfattah:jocs:2017}, LU~\cite{ahtd17}, and QR~\cite{atd22} factorizations. Unlike BLAS and one-sided factorizations which require a fixed number of FLOPs regardless of the matrix numerical properties, all SVD algorithms are iterative, and the matrices in the batch may require a different number of iterations even when they all have same size.
This adds a layer of complexity to the design of efficient batch SVD algorithms.
Dong et al.~\cite{dhtd14} were the first to propose an algorithm to the batch bi-diagonalization phase~\ref{it:phase1}, but not the subsequent diagonalization~\ref{it:phase2}.
Badolato et al.~\cite{bpf15} implemented batch variants of the one-sided Jacobi SVD algorithm and of the bi-diagonalization approach of Demmel and Kahan~\cite{deka90}.
They compared these two implementations on an image processing application, concluding that on a GPU their Jacobi-type implementation is more efficient than their bi-diagonalization one when processing a batch of small matrices.
Boukaram et al.~\cite{btlk18} developed a batched block Jacobi SVD algorithm.
This algorithm, which is part of the KBLAS library~\cite{akl16}, applies the blocked Jacobi rotations in each matrix serially, and it can only saturate the GPU if the matrices are of order 64 or less. The developed solver in KBLAS lacks support for complex matrices, and does not compute the right singular vectors. In fact, the KBLAS solver would require a separate batch linear solver to recover the right singular vectors (solving for $V$: $AV = U\Sigma$).
The most efficient batch SVD implementation of which we are aware is due to Huang et al.~\cite{hylz22}, which is called Wcycle-SVD. They propose a parallel blocked Jacobi SVD algorithm that applies the block Jacobi rotations to the Gram matrix in parallel. However, the implementation is available only in real double precision, and sometimes fails to successfully complete the decomposition.


\section{Algorithmic Background}
\label{sec:algo}
\newcommand{\Ak}[1][k]{\ensuremath{A^{(#1)}}}       
\newcommand{\Jl}[3]{\ensuremath{J(#1, #2, #3)}}     
\newcommand{\Js}[3]{\ensuremath{\widetilde J(#1, #2, #3)}} 
\newcommand{\Jb}[4]{\ensuremath{\widetilde J(#1, #2, #3)_{#4}}} 

\paragraph{Jacobi Hermitian eigensolver}
The Jacobi method predates digital computing. It was first proposed by Jacobi~\cite{jaco46} in 1846, but it only became widely used with the advent of computers in the 1950s.
Given $A \in \Cnn$, the method applies a sequence of plane rotations to compute the eigendcomposition
\begin{equation}
    \label{eq:eigendecomposition}
    M^H A M = \Lambda,
\end{equation}
where $M \in \Cnn$ is Hermitian and $\Lambda \in \Cnn$ is diagonal.
Starting with $\Ak[0] = A$, at step $k$ the algorithm computes a Jacobi matrix $\Jl{i}{j}{k}$ such that
\begin{equation}\label{eq:transform}
    \Ak[k+1] = J^H \Ak J,
\end{equation}
is closer than $\Ak$ to being diagonal.
We will make this statement precise after explaining how the matrix $\Jl{i}{j}{k}$ is computed from $\Ak$.

Let $a_{p,q}$ denote the element in position $(p,q)$ of $\Ak$.
$A$ is Hermitian, and the transformations in~\eqref{eq:transform} preserve this property, thus $\Ak$ is also Hermitian with $a_{q,p} = \overline{a_{p,q}}$.
After selecting a pair of column indices $(i,j)$, with $1 \le i < j \le n$, the algorithm needs to compute a complex plane rotation in the $(i,j)$-plane to annihilate $a_{i,j}$ and $a_{j,i}$. This can be achieved by solving the $2 \times 2$ Hermitian eigenvalue problem
\begin{equation}\label{eq:background-eigenproblem-jacobi}
\begin{bmatrix}
\cos\theta & -e^{i\phi}\sin\theta \\
e^{-i\phi}\sin\theta & \cos\theta
\end{bmatrix}^H
\begin{bmatrix}
a_{i,i} &  \overline{a_{j,i}}\\
a_{j,i} &  a_{j,j}\\
\end{bmatrix}
\begin{bmatrix}
\cos\theta & -e^{i\phi}\sin\theta \\
e^{-i\phi}\sin\theta & \cos\theta
\end{bmatrix}
=
\begin{bmatrix}
\lambda_{i} &  0\\
0 &  \lambda_{j}\\
\end{bmatrix}.
\end{equation}
The eigenvectors of~\eqref{eq:background-eigenproblem-jacobi} can be found by setting
\begin{equation}\label{eq:sin-cos}
  \phi = \arg(a_{i,j}),\quad
  \cos \theta = \frac{1}{\sqrt{1 + t_{\min}^{2}}}, \qquad
  \sin \theta = 
  \frac{t_{\min}}{\sqrt{1 + t_{\min}^{2}}},\qquad
  t_{\min} = \frac{\sign(\tau)}{\abs{\tau} + \sqrt{1 + \tau^{2}}},
  \qquad
  \tau = \frac{a_{i,i} - a_{j,j}}{2 \abs{a_{i,j}}},
\end{equation}
if $a_{i,j} \neq 0$, and $\cos \theta = 1$ and $\sin \theta = 0$, otherwise.
Embedding this $2 \times 2$ rotation into the identity matrix of size $n$ at rows and columns $i$ and $j$ yields
\begin{equation}\label{eq:jacobi-rotation}
    \Jl{i}{j}{k} =
    \begin{bmatrix}
        I_{i-1} & & & & & \\
        & \cos\theta & &-e^{-i\phi}\sin\theta & & \\
        & & I_{j-i-1} & & & \\
        & e^{-i\phi}\sin\theta & & \cos\theta & &\\
        & & & & I_{n-j} \\
    \end{bmatrix},
\end{equation}
which completes the step.
A series of steps that annihilates all non-diagonal elements of $A$ in turn is called a \emph{sweep}. The pseudocode of this method is given in~\cref{alg:jacobi_unblocked}, which requires an accuracy threshold to decide whether an off-diagonal element should be annihilated. Such a threshold is set to the product $ku$, where $k$ is a tolerance parameter set by the user, and $u$ is the unit roundoff $u = 2^{-s}$, where $s$ is the number of significand bits, including the implicit bit.
\begin{algorithm}[t]
\KwData{Hermitian matrix $A \in \Cnn$.}
\KwParam{Maximum number of sweeps \tt{max\_nsweeps}, tolerance threshold $k$, and unit roundoff $u$.}
\KwResult{The eigendecomposition \eqref{eq:eigendecomposition} of $A$.}
$V \gets I$\;
$converged \gets \false$\;
$sweeps \gets 0$\;
\While{\mnot{} $converged$ \mand{} sweeps $<$ \tt{max\_nsweeps}}{
\tcp{Begin Jacobi sweep.}
$converged \gets\true$\;
$sweeps \gets sweeps + 1$\;
\For{each pair $(i, j)$ with $1 \le i < j \le n$} {
\If{$ \left | a_{j,i} \right | \geq k u \sqrt{a_{i,i} a_{j,j}}$ } {
$converged \gets \false$\;
Compute $\phi$, $\cos \theta$, and $\sin \theta$ in~\eqref{eq:sin-cos}.\;
Generate the matrix $\Jl{i}{j}{k}$ in \eqref{eq:jacobi-rotation}.\;
$A \gets \Jl{i}{j}{k}^H\; A \;\Jl{i}{j}{k}$\;
}}}
\caption{The unblocked Jacobi Hermitian eigensolver}
\label{alg:jacobi_unblocked}
\end{algorithm}

The analysis of the convergence of the Jacobi algorithm relies on the off-norm operator, which if $A \in \Cnn$ is Hermitian is defined as
\begin{equation*}
\off(A) = \sqrt{2\sum_{p=1}^{n}\sum_{q=p+1}^{n} \abs{a_{p,q}}^{2}} = \fnorm{A - \diag(A)},
\end{equation*}
where $\diag(A)$ returns the diagonal matrix $D \in \Cnn$ with $d_{pp} = a_{pp}$.
It is easy to show that
\begin{equation}\label{eq:scalar-off-norm}
    \off\bigl(\Ak[k+1]\bigr)
    = \off\bigl(\Ak\bigr)^{2} - 2 \abs{a_{i,j}}^{2}.
\end{equation}
An important consequence of~\eqref{eq:scalar-off-norm} is that if $a_{i,j} \neq 0$, then every step of the Jacobi algorithm reduces the off-diagonal norm.

A Jacobi method is globally convergent if, for any $A \in \Cnn$, we have
\begin{equation*}
    \lim_{k \to \infty} \Ak = \Lambda,
\end{equation*}
where $\Lambda \in \Cnn$ is diagonal. As each step of the algorithm is a similarity transformation, the values along the diagonal of $\Lambda$ are the eigenvalues of $A$, and an eigendecomposition can be obtained by noticing that
\begin{equation*}
M^H A M = \Ak,\qquad
M = \prod_{\ell=0}^{k}J_\ell
\end{equation*}

The convergence of the Jacobi SVD algorithm is affected by the order in which the off-diagonal elements are annihilated.
Jacobi showed that the method converges if, at each step, one annihilates a \emph{pivot}---an element $a_{i,j}$ such that $\abs{a_{i,j}} = \max_{p \neq q} \abs{a_{p,q}}$.
\emph{Pivoting} enables faster convergence but, in practice, it is rarely used, because finding the element of largest magnitude requires a search of quadratic complexity, which has a severe impact on performance.
Practical implementation typically rely on \emph{cyclic strategies}, in which the Jacobi method applies rotations to all off-diagonal index pairs in a fixed, predetermined order that is repeated sweep after sweep.
Forsythe and Henrici~\cite{fohe60} showed that the method converges globally for the row- and column-cyclic ordering, where one proceeds one row or one column at a time, respectively, annihilating each non-diagonal element of $A$ in turn.
Hansen~\cite{hans63} provided the first examples of cyclic strategies that do not converge and introduced the notion of equivalence of orderings, where commuting pivots can be swapped without changing convergence behavior.
Row- and column-cyclic orderings are in the same equivalence class, and any cyclic ordering in that class is globally convergent.
Nazareth~\cite{naza75} further extended these results to a broader constructive class of cyclic orderings.
Hari and Begović Kovač~\cite{habe16} introduced the class of \emph{generalized serial cyclic strategies}.
They proved that each such ordering yields a uniform per-sweep contraction, and hence global convergence of the Jacobi method.
For the special case of $n = 4$, one can further prove that the Jacobi method is globally convergent for all 720 cyclic orderings, although a uniform contraction is only achieved after three sweeps~\cite{beha17a}, in general.

\begin{figure}[t]
\centering
\includegraphics[width=0.8\linewidth]{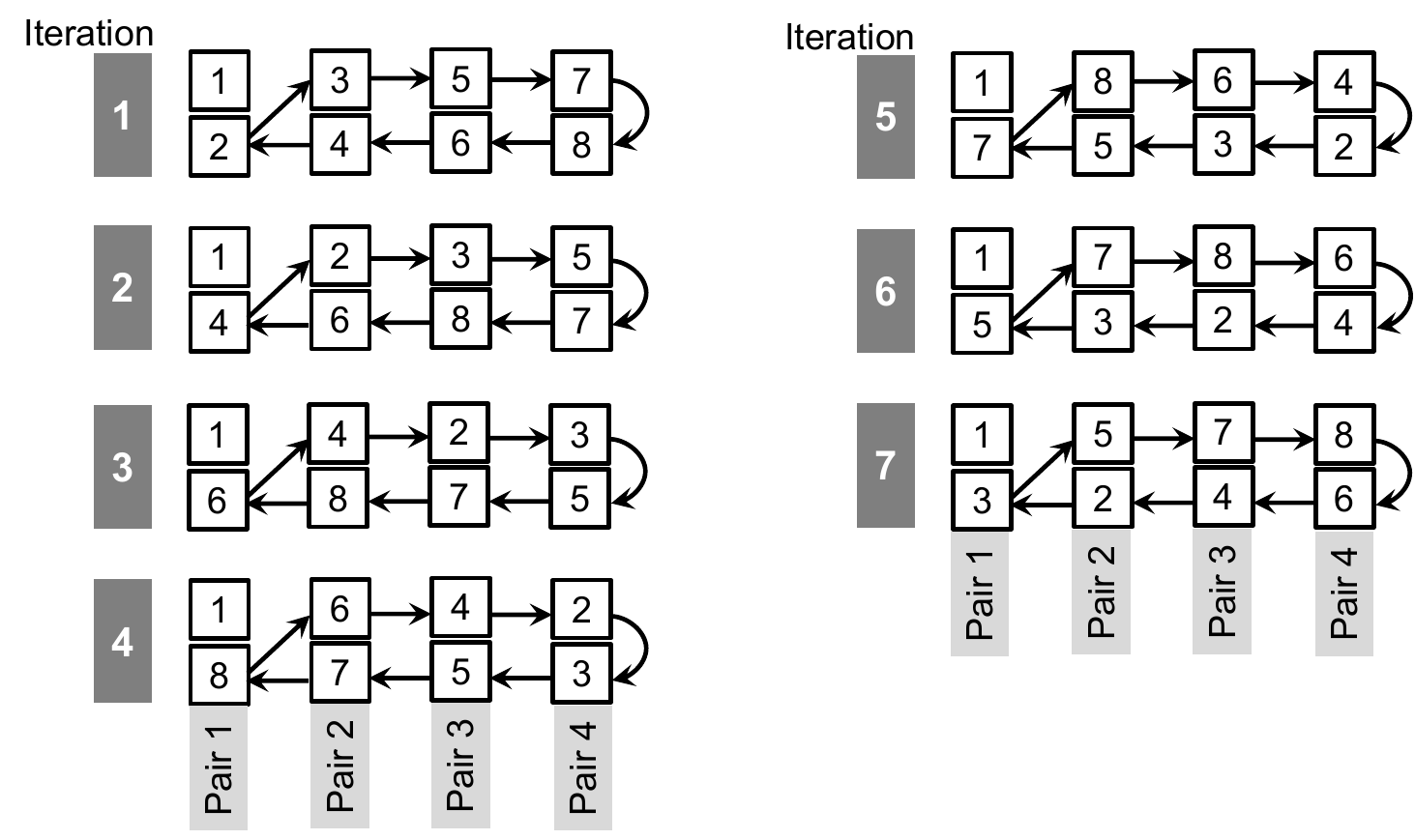}
\caption{Example of the parallel ordering with eight block columns. Each Jacobi sweep consists of seven iterations. Each iteration contains four disjoint pairs of block-columns.}
\Description{An illustration of the round-robin parallel ordering approach.}
\label{fig:rr}
\end{figure}

One of the advantages of the Jacobi method is the high potential for parallel execution.
In fact, off-diagonal elements corresponding to disjoint ($i, j$) pairs can be annihilated concurrently.
Following the seminal work of Sameh~\cite{same71}, many parallel Jacobi algorithms have been proposed in the literature, and orderings suitable for parallel execution are well understood.
Brent and Luk~\cite{brlu85} introduced the round-robin ordering, which we use in this paper; an example of this ordering for $n = 8$ is depicted in~\cref{fig:rr}.
Luk and Park~\cite{lupa89} described five parallel orderings that are in the class of provably convergent orderings of Hansen, and Shroff and Schreiber~\cite{shsc89,shsc91} characterized all parallel orderings that are equivalent to the row-cyclic ordering.
Other parallel orderings were introduced in the literature, such as, for example ring and odd-even orderings~\cite{zhbr95}.

A second advantage of the Jacobi method is that it may be significantly more accurate than bi-diagonalization algorithms.
Demmel and Veselić~\cite{deve92} prove that if $A$ is positive definite, then the Jacobi eigensolver is optimally accurate up to factors that depend on the size of the matrix, and Mathias~\cite{math95} generalizes this result.
Mascarenhas~\cite{masc94} investigates, mostly experimentally, the case of general Hermitian matrices.

\paragraph{One-sided Jacobi SVD algorithms}
There are two standard routes to compute the SVD using Jacobi-type algorithms.
The two-sided Jacobi SVD algorithm, first proposed by Kogbetliantz~\cite{kogb55}, is a direct extension of the Hermitian Jacobi algorithm to general matrices, which applies right and left rotations to $A$ to annihilate off-diagonal matrix entries. The algorithm requires partitioning the matrix across both rows and columns. It can be applied on rectangular matrices after a QR factorization as a pre-processing step~\cite{dongarra2018singular}.

Here, we focus on the one-sided Jacobi SVD algorithm, due to Hestenes~\cite{hest58}, which applies the Jacobi eigensolver to the Hermitian matrix $A^HA$. The one-sided algorithm does not compute $A^HA$ explicitly.
Instead, it computes the Jacobi rotation $\Jl{i}{j}{k}$ that annihilates an element of $A^HA$ and it applies it implicitly by producing the sequence of matrices
\begin{equation*}
    \Ak[0] = A,\qquad \Ak[k+1] = \Ak \Jl{i}{j}{k}.
\end{equation*}
Compared to the two-sided Jacobi SVD approach, the one-sided Jacobi algorithm is simpler to implement on parallel architectures. It requires partitioning the input matrix across columns only, and can be directly applied to rectangular matrices without a QR factorization pre-processing. We now describe the method that operates on the right Gram matrix $A^HA$, but the algorithm could be derived analogously by working on the left Gram matrix $AA^H$ and applying transformations to the left of $A$.

At each step, the algorithm selects a pair of column indices $(i,j)$ with $1 \le i < j \le n$,
and computes the Jacobi rotation that annihilates the element in position $(i,j)$ of $A^HA$.
To compute this rotation, the algorithm solves the $2 \times 2$ eigenvalue problem
\begin{equation}\label{eq:background-svd-jacobi}
\begin{bmatrix}
\cos\theta & -e^{i\phi}\sin\theta \\
e^{-i\phi}\sin\theta & \cos\theta
\end{bmatrix}^H
\begin{bmatrix}
\acol{i}^H\acol{i} & \acol{i}^H\acol{j} \\
\acol{j}^H\acol{i} & \acol{j}^H\acol{j}
\end{bmatrix}
\begin{bmatrix}
\cos\theta & -e^{i\phi}\sin\theta \\
e^{-i\phi}\sin\theta & \cos\theta
\end{bmatrix}
=
\begin{bmatrix}
\sigma_{i} &  0\\
0 &  \sigma_{j}\\
\end{bmatrix},
\end{equation}
where $\acol{p}$ represents the $p$th column of $A$.
The eigenproblem~\eqref{eq:background-svd-jacobi} is analogous to~\eqref{eq:background-eigenproblem-jacobi} and has the same solution~\eqref{eq:sin-cos}.
Embedding this rotation into the $n \times n$ identity yields the Jacobi rotation~\eqref{eq:jacobi-rotation}, and this concludes the step.
The pseudocode of the method is given in \cref{alg:gesvj_unblocked}.

\newlength{\algwidthone}
\newlength{\algwidthtwo}
\setlength{\algwidthone}{\widthof{$g_{j,j}$}}
\setlength{\algwidthtwo}{\widthof{$\acol{i}\;\;\acol{j}$}}
\begin{algorithm}[t]
\KwData{General Matrix $A \in \Cmn$ with $m \ge n$.}
\KwParam{Maximum number of sweeps \tt{max\_nsweeps}, tolerance threshold $k$, and unit roundoff $u$.}
\KwResult{The reduced SVD~\eqref{eq:svd} of $A$.}
$V \gets I$\;
$converged \gets \false$\;
$sweeps \gets 0$\;
\While{\mnot{} $converged$ \mand{} sweeps $<$ \tt{max\_nsweeps}}{
\tcp{Begin Jacobi sweep.}
$converged \gets\true$\;
$sweeps \gets sweeps + 1$\;
\For{each pair $(i, j)$ with $1 \le i < j \le n$} {
$\makebox[\algwidthone][l]{$g_{i,i}$} \gets \acol{i}^H \acol{i}$\;
$\makebox[\algwidthone][l]{$g_{j,j}$} \gets \acol{j}^H \acol{j}$\;
$\makebox[\algwidthone][l]{$g_{j,i}$} \gets \acol{j}^H \acol{i}$\;
\If{$ \left | g_{j,i}\right | \geq k u \sqrt{g_{i,i} g_{j,j}}$ } {
$converged \gets \false$\;
Find $\Js{i}{j}{k}$ such that
$\Js{i}{j}{k}^H\begin{bmatrix}
g_{i,i} & \overline{g_{j,i}} \\
g_{j,i} & g_{j,j} \\
\end{bmatrix}\Js{i}{j}{k}$ is diagonal.\;
$[\makebox[\algwidthtwo][c]{\acol{i}\;\;\acol{j}}] \gets
[\makebox[\algwidthtwo][c]{\acol{i}\;\;\acol{j}}]\;\Js{i}{j}{k}$\;
$[\makebox[\algwidthtwo][c]{\vcol{i}\;\;\vcol{j}}] \gets
[\makebox[\algwidthtwo][c]{\vcol{i}\;\;\vcol{j}}]\;\Js{i}{j}{k}$\;
}}}
\tcp{Compute singular values and normalize left singular vectors.}
\For{$i$ from 1 to $n$} {
$\sigma_{i} \gets \lVert \acol{i} \rVert_2$\;
$U_i \gets \acol{i} / \sigma_i$\;
}
\tcp{Sort $\Sigma$ in descending order and reorder $U$ and $V$ accordingly.}
\caption{The unblocked one-sided Jacobi SVD algorithm}
\label{alg:gesvj_unblocked}
\end{algorithm}

One can show that
\begin{equation}\label{eq:block-off-norm}
  \off\big((\Ak[k+1])^{H}\Ak[k+1]\big)^{2}
  = \off\big((\Ak)^{H}\Ak\big)^{2} - 2 \abs*{(\acol{i}^{(k)})^H\acol{j}^{(k)}}^{2},
\end{equation}
where $\acol{p}^{(k)}$ denotes the $p$th column of $\Ak$.
This is analogous to~\eqref{eq:scalar-off-norm}.
Therefore, unless the $i$th and $j$th column of $\Ak$ are orthogonal, every step of the Jacobi algorithm reduces the off-diagonal norm of $(\Ak[k+1])^{H}\Ak[k+1]$.

For a given cyclic ordering, the one-sided Jacobi method is globally convergent if it converges to a matrix with orthogonal columns for any $A \in \Cnn$.
All convergence results follow by applying the Hermitian result to the Gram matrix $A^HA$.
In particular, any ordering that guarantees global convergence of the Hermitian eigensolver will also guarantee convergence of the one-sided SVD algorithm, and all previously discussed results apply.

The accuracy of the algorithm can be improved, in many cases, by preconditioning the eigensolver.
Demmel et al.~\cite{dges99} show that, if preconditioned with a rank-revealing decomposition, the one-sided Jacobi SVD algorithm can compute singular values with high relative accuracy even when these have widely varying magnitudes.

\vspace{-5pt}\paragraph{Efficient implementations}
In the literature, efforts to improve the performance of the Jacobi methods have followed two directions. On the one hand, authors have tried to reduce the number of FLOPs the algorithm requires and improve its convergence speed.
Veselić and Hari~\cite{veha89} show that the efficiency of the Jacobi Hermitian eigensolver can be improved by applying the one-sided SVD algorithm to the pivoted Cholesky factor of the Hermitian positive definite matrix.
Inspired by this, Drmač and Veselić~\cite{drve08I} suggest running the one-sided Jacobi SVD algorithm on the triangular factor of a column pivoted QR factorization of $A$, which is equivalent to running the Jacobi method on the triangular factor of a pivoted Cholesky factorization of the Gram matrix $A^HA$.
To further improve the efficiency of this method, the same authors~\cite{drve08II} study how this preconditioned method can be implemented effectively.

On the other hand, authors have looked at efficient ways of exploiting the memory hierarchy through \emph{blocking}.
On modern hardware, the performance of linear algebra algorithms is often limited by data movement across the memory hierarchy, not by the arithmetic throughput of the processing unit.
Processing data by blocks, rather than at the element level, is a customary technique to improve the performance of these algorithms, as it improves data locality and cache reuse.
To the best of our knowledge, Van Loan~\cite{van86} was the first to propose a block Jacobi method, focusing on the case of the two-sided Jacobi algorithm.
This algorithm builds on previous work by Brent and Luk~\cite{brlu85} and Brent, Luk, and Van Loan~\cite{blv85}.
Foulser~\cite{foul89} proposed an algorithm based on a different pivoting strategy, similar to the original maximum pivoting used by Jacobi.
Pivoting enables faster convergence but, in practice, selecting the largest off-diagonal entry (or block) at every step is too expensive, which motivates the use of cheaper predetermined cyclic strategies.
For block algorithms, the literature has also considers dynamic ordering schemes, in which the pairs of indices to annihilate are selected adaptively according to the current state of the matrix rather than a fixed sweep pattern.
In the context of block-Jacobi SVD, this idea was introduced by Bečka, Okša, and Vajteršic~\cite{bov02}, who proposed choosing block pairs so as to maximize the reduction of the off-diagonal Frobenius norm at each step, formulating the selection problem in terms of weighted matching on the current block graph.
Subsequent work extended dynamic ordering to the parallel one-sided block-Jacobi SVD algorithm~\cite{bov15} and developed several practical variants with different computational and communication trade-offs.
More recently, the convergence of the dynamically-ordered two-sided block-Jacobi SVD algorithm to compute singular triplets has been analyzed in detail~\cite{oyv22}.

We now recall the block one-sided Jacobi SVD method using a cyclic block strategy, and discuss existing results on the convergence of this method.
Let $\nb$ be the block size, and let $\ell = n / \nb$.
We can assume without restriction that $n$ is a multiple of $\nb$, as the matrix can be padded with columns of zero if this is not the case.
Partitioning the matrix as
\begin{equation*}
  A = [\abcol{1} \; \abcol{2} \; \cdots \; \abcol{\ell}], \qquad \abcol{j} \in \C^{m \times \nb},
\end{equation*}
induces a conformal block structure on $A^HA$, which is a square matrix with $\ell$ rows and $\ell$ columns of $\nb \times \nb$ blocks.
A Jacobi step selects a pair of block indices $(i,j)$ and solves the $2\nb \times 2\nb$ Hermitian eigenvalue problem
\begin{equation}
  \label{eq:block-gram}
  \Js{i}{j}{k}^H G_{i,j} \Js{i}{j}{k} = \Lambda
  ,\qquad
  \Js{i}{j}{k} = \begin{bmatrix}
    \Jb{i}{j}{k}{1,1} & \Jb{i}{j}{k}{1,2}\\
    \Jb{i}{j}{k}{2,1} & \Jb{i}{j}{k}{2,2}
  \end{bmatrix},\qquad
  G_{i,j} = [A_i\;\;A_j]^H[A_i\;\;A_j] = \begin{bmatrix}
  \abcol{i}^H\abcol{i} & \abcol{i}^H\abcol{j} \\
  \abcol{j}^H\abcol{i} & \abcol{j}^H\abcol{j}
  \end{bmatrix},
\end{equation}
where $\Lambda \in \Ctnbtnb$ is diagonal.
The matrix $\Js{i}{j}{k}$ is embedded into the unitary matrix
\begin{equation}\label{eq:right-sv}
\Jl{i}{j}{k} =
\begin{bmatrix}
  I_{(i-1)\nb} &         & & & \\
                           & \Jb{i}{j}{k}{1,1} & & \Jb{i}{j}{k}{1,2} & & \\
                           &         & I_{(j-i-1)\nb} &        & \\
                           & \Jb{i}{j}{k}{2,1} &                            & \Jb{i}{j}{k}{2,2} & \\
                           &         &                            &        &
  I_{(\ell - j)\nb}
\end{bmatrix} \in \Cnn,
\end{equation}
and, as in previous cases, the update $\Ak[k+1] = \Ak \Jl{i}{j}{k}$ completes the step.

Drmač~\cite{drma09} shows that it is not necessary to solve the eigenvalue problem~\eqref{eq:block-gram} exactly: the one-sided Jacobi SVD method with row- or column-cyclic pivoting is globally convergent, even when the matrix has repeated or clustered singular values, as long as each $\Js{i}{j}{k}$ satisfies the following two properties.
\begin{enumerate}[label=P\arabic*., ref=P\arabic*]
  \item $1 \ge \sigma_{\min}(J_{1,1}) = \sigma_{\min}(J_{2,2}) \ge \gamma$, for some positive constant $\gamma$ that depends only on the size of the diagonal blocks. This property is called the \emph{uniformly bounded cosines} (UBC) property after \cite[Def.~2.4]{drma09}. \label{it:bVij-1}
  \item $\off\big(\Js{i}{j}{k}^H G_{i,j} \Js{i}{j}{k}\big) \le \rho \off(G_{i,j})$, for $0 \le \rho < 1$ independent of $k$. In other words, the transformation must reduce the off-norm of the block $G_{i,j}$. For $\rho = 0$, we obtain that $\Js{i}{j}{k}$ contains the eigenvectors of $G_{i,j}$. \label{it:bVij-2}
\end{enumerate}
These results were extended by Hari~\cite{hari21} to the generalized eigenvalue problem and to the larger class of generalized serial orderings.
Hari and Begović Kovač~\cite{beha17} established the convergence of both cyclic and quasi-cyclic block Jacobi methods under very general assumptions on the block partition and the pivot structure, thereby unifying and extending earlier results.
The structural conditions required for convergence of the generalized serial pivot orderings in the real case can be transferred---appropriately modified---to the complex setting~\cite{beha24}.


\section{Experimental Setup}
\label{sec:setup}
All the experiments in this paper are conducted on two systems.
\begin{itemize}[nolistsep]

    \item An NVIDIA Grace--Hopper GH200 system, which features a 72-core NVIDIA Grace CPU (ARM-based Neoverse-V2) with a maximum clock rate of 3.438 GHz (frequency scaling at 92\%). The system hosts an NVIDIA H200 GPU with 132 streaming multiprocessors, 96 GB of memory, and a maximum clock rate of 1.98 GHz. On this system, we use a development branch of the MAGMA library, compiled with CUDA 12.8.1, and the NVIDIA performance libraries (NVPL) version 0.2 for testing purposes.
    
    \item An AMD system with four MI300A Accelerated Processing Units (APUs). Each APU features a 24-core AMD EPYC CPU (x86 Zen 4 architecture), with a maximum clock rate of 3.7 GHz (frequency scaling at 45\%), and a GPU with 228 Compute Units (CUs), 128 GB of memory, and a maximum clock rate of 2.1 GHz. On this system, we use a development branch of the MAGMA library, compiled with ROCm 7.0.1, and Intel oneMKL library version 2025.0.2.

\end{itemize}
\section{Initial Design}
\label{sec:initial_design}

\newcommand{\GEMM}{\texttt{GEMM}}
\newcommand{\HERK}{\texttt{HERK}}

This section describes the initial design of the batch SVD solver that processes simultaneously \texttt{batch} matrices.
This design represents our \emph{baseline implementation}.
The following sections introduce several layers of optimizations to improve the time-to-solution. For every design optimization, we will show the corresponding gain in performance and, if needed, prove that the accuracy of the algorithm is not affected.

The initial design is blocked, meaning that an input matrix $A \in \Cmn$ is split into block columns of a chosen width $nb$. The number of block columns is $\bc = \left \lceil \frac{n}{\nb}\right \rceil$, and the corresponding Hermitian matrix $A^HA$ consists of $\bc \times \bc$ square blocks of order $\nb$. For simplicity, the description of the algorithm assumes that $\nb$ fully divides $n$ and that $\bc$ is even. During a Jacobi sweep, the algorithm chooses $\bc/2$ disjoint pairs of block columns. For each pair $A_{i}$ and $A_{j}$, the algorithm solves the eigenproblem~\eqref{eq:block-gram} and then post-multiplies $A$ with the Jacobi matrix containing the eigenvectors.
The post multiplications affect only the block columns $\abcol{i}$ and $\abcol{j}$. A single sweep of the block Jacobi method is concluded when all possible pairs of block columns have been processed. The algorithm executes multiple sweeps, cycling through all possible block pairs, until convergence. Each sweep involves $\frac{1}{2}\bc(\bc-1)$ pairs, which can be distributed across $(\bc-1)$ iterations of $\frac{1}{2}\bc$ pairs each.
Each of these iterations comprises the following steps.
\begin{enumerate}[nolistsep,label=S\arabic*.,ref=S\arabic*]
\item Generate a \emph{parallel ordering}, which is a list of $\frac{1}{2}\bc$ independent block-column pairs.\label{it:s1}
\item Compute the $\frac{1}{2}\bc$ Gram matrices.\label{it:s2}
\item Solve the Hermitian eigenvalue problem of the Gram matrices.\label{it:s3}
\item Update the left singular vectors (always).\label{it:s4}
\item Update the right singular vectors (if required).\label{it:s5}
\end{enumerate}
The left singular vectors are updated even when the user does not require them, because successively applying the eigenvectors in \ref{it:s3} to $A$ produces the matrix $A^{(k)}$, which converges to $U\Sigma$.
As long as $\bc > 2$, all steps except \ref{it:s1} can be performed using batch routines, even when the SVD of only one matrix is being computed.

\subsection{Parallel Ordering}
\label{subsec:rr}
Our algorithm relies on the round-robin parallel ordering introduced in \cref{sec:algo}, which is non-pivoted. The SVD solver we propose is not tied to a particular ordering scheme, and, in our software design, the index-pair generation is entirely decoupled from the remaining computational steps.
The index pairs are generated using a GPU kernel. Once generated, the same kernel uses the index pairs to set up several pointer arrays that correspond to the batch operations carried out in the subsequent computational stages. If another ordering is to be used, another GPU kernel must be developed for this step. The other components of the solver will not change.

\subsection{Computation of the Gram Matrices}
\label{subsec:gram}
The Gram matrix $G_{i,j}$ in \eqref{eq:block-gram} can be computed using a Hermitian rank-$k$ update, which is a standard operation (\HERK) in the BLAS~\cite{blas}. When present in vendors' libraries, batch BLAS kernels are typically highly optimized and ensure performance portability across different architectures. However, the batch \HERK{} kernel is only available in rocBLAS~\cite{rocblas} but not in cuBLAS~\cite{cublas}.

Moreover, in our particular use case, using the batch \HERK{} kernels would require the explicit formation of the matrix $[\abcol{i}\;\;\abcol{j}]$, which will incur a data movement overhead unless the two block columns $\abcol{i}$ and $\abcol{j}$ are adjacent in memory.

\begin{figure}[!htb]
\centering
\includegraphics[width=0.4\linewidth]{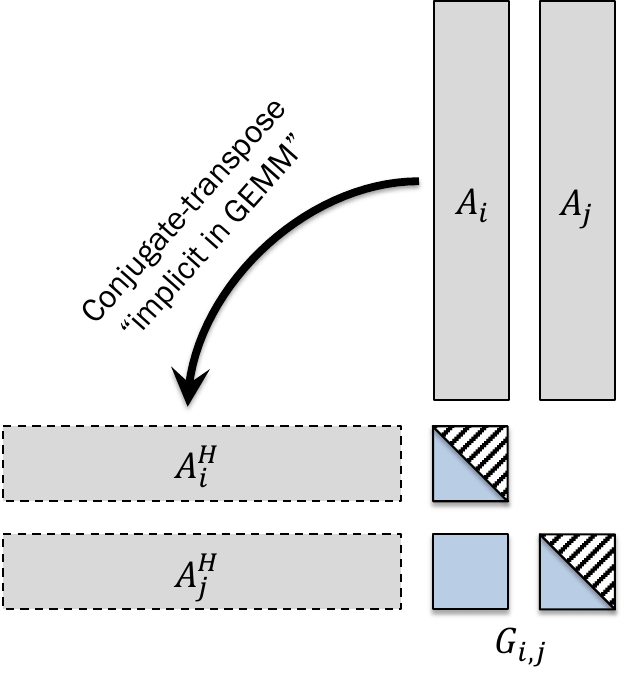}
\caption{Computing the Gram matrix using three concurrent matrix multiplications}
\label{fig:gram}
\end{figure}

Therefore, we prefer to use batch \GEMM{} in our implementation. Since $G_{i,j}$ is Hermitian, we only need to compute the lower or upper triangular part. This can be performed using three concurrent \GEMM{} calls, each of size $\nb \times \nb$, as shown in \cref{fig:gram}. Given a block-column pair $(i,j)$, the three concurrent \GEMM's are $\abcol{i}^HA_{i}$, $\abcol{j}^HA_{i}$, and $\abcol{j}^H A_{j}$, each producing a sub-block of the Gram matrix of size $\nb \times \nb$. The use of a batch \GEMM{} implies $25\%$ overhead, as the dashed areas in~\cref{fig:gram} are computed unnecessarily. We believe that this overhead is negligible, as long as $nb$ is sufficiently small.
Each Jacobi SVD sweep requires the computation of $\frac{1}{2}\bc$ Gram matrices, and the whole solver will require $\frac{3}{2}\bc \cdot \tt{batch}$ matrix multiplications, each producing an $\nb \times \nb$ output product of an $\nb \times m$ and an $m \times \nb$ term.

The \GEMM s in \cref{fig:gram} have a relatively large inner dimension, which corresponds to the number of rows of $A$, compared with the size of the small Gram matrix. Such an unconventional shape could be poorly optimized in vendors' BLAS library, thus we investigate which library provides the most efficient \GEMM{} kernel for this shape.

\begin{figure}[!htb]
\centering
\includegraphics[width=0.49\linewidth]{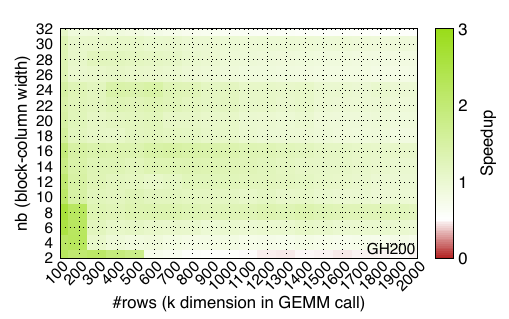}
\includegraphics[width=0.49\linewidth]{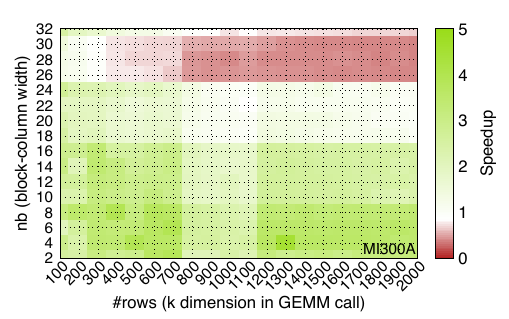}
\caption{Heatmaps showing the speedups for MAGMA's own batch GEMM kernel over the vendor's BLAS library for the use case of computing relatively small Gram matrices. Results are shown for double precision on the GH200 system (left) and the MI300A APU (right). Green color (speedups $>1$) means that using MAGMA is recommended for the given dimensions. Red color (speedups $<1$) means that the vendor BLAS is used.}
\label{fig:gemm_cn_speedup}
\end{figure}

The MAGMA library~\cite{abcg24} provides an open source batch \GEMM{} kernel that is highly parameterized and tunable~\cite{abdelfattah2016performance}. While it cannot match the asymptotic performance of the vendors' own BLAS libraries, it can be instantiated to target specific unconventional shapes such as those in \cref{fig:gram}. We conducted an offline tuning experiment for $\nb \in [2:32]$ and for $m \leq 2000$, comparing the performance of the cuBLAS and rocBLAS to MAGMA on all four BLAS data types. \Cref{fig:gemm_cn_speedup} reports the results for the ``D'' variant, which corresponds to the real double precision implementation. The figure shows that MAGMA's batch \GEMM{} kernel can outperform the vendor's own implementation in most cases, with speedups up to $3\times$ on the GH200 system and up to $5\times$ on the MI300A APU.

Therefore, we developed a thin decision layer that selects the best performing kernel based on the results of this offline tuning experiment. If the dimensions are outside the range of the data collected, the implementation defaults to the vendor implementation.
\subsection{Hermitian Eigenvalue Problem}
\label{subsec:heevj}

A critical component of the blocked SVD solver is the solution of the Hermitian eigenvalue problem~\eqref{eq:block-gram}, where $\Js{i}{j}{k}$ is the eigenvectors used to update the block-column pair $(i,j)$ in the matrix $A^{(k)}$ as well as the right singular vectors.
In practice, any batch Hermitian eigensolver could be used, including those provided by the cuSOLVER and the rocSOLVER libraries. However, we prefer to use our own implementation of the two-sided Jacobi method in \cref{alg:jacobi_unblocked}; this will allow us to control some parameters of the solver that are inaccessible through the interface provided by vendor libraries.

The batch eigensolver implements a parallel version of \cref{alg:jacobi_unblocked} that uses the round-robin parallel ordering. An important requirement for the eigensolver is that the matrix and its eigenvectors should be small enough to fit in the shared memory of the GPU. This requirement limits the blocking size $\nb$ used in the batch SVD solver, but the limit is acceptable in most cases, as we now explain. The Gram matrix for a single block-column pair is of order $2\nb$, solving the eigenvalue problem (with vectors) requires $8\nb^2$ elements. A lower bound on the number of elements is given by taking the largest element size on the smallest shared memory. This is obtained by using the double-complex data type, which requires 16 bytes (128 bits) per element, and the shared memory of the AMD APU, which is 65KiB; this limits $\nb$ to a maximum of $22$.
\begin{figure}[!htb]
\centering
\includegraphics[width=0.49\linewidth]{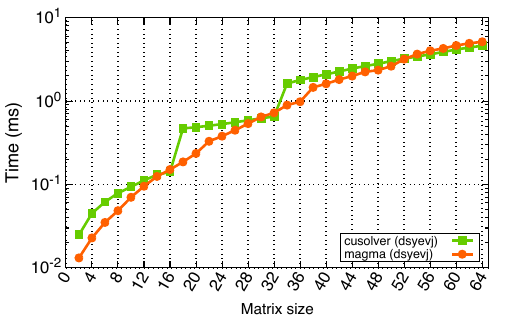}
\includegraphics[width=0.49\linewidth]{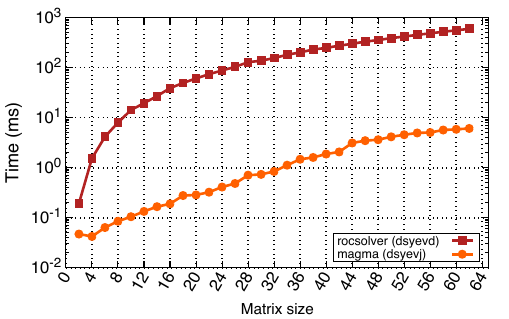}
\caption{Time-to-solution of MAGMA's batch Hermitian eigensolver, with vectors computed. Results are shown for $1000$ double-precision matrices, on a GH200 system (left) and on an MI300A APU (right).}
\label{fig:heevj_perf}
\end{figure}

In terms of performance, the developed eigensolver is either comparable with or superior to the vendor's own implementation. \Cref{fig:heevj_perf} compares the execution time of the MAGMA solver with that of cuSOLVER, on the GH200 system, and of rocSOLVER, on the MI300A system. We can use the LAPACK naming convention to deduce that the cuSOLVER routine (\tt{dsyevj}) uses a Jacobi based algorithm, while the rocSOLVER routine (\tt{dsyevd}) uses an algorithm based on reduction to tridiagonal form followed by a divide-and-conquer algorithm, when vectors are required. The figure shows that, on the GH200 system, cuSOLVER
is at best $12\%$ faster than MAGMA for sizes larger than $52$, but the latter can achieve speedups up to $2.5\times$ on smaller matrices. On the MI300A APU, the MAGMA solver is significantly faster than the rocSOLVER implementation, with speedups between $4.2\times$ and $228.8\times$.
\begin{figure}[!htb]
\centering
\includegraphics[width=0.4\linewidth]{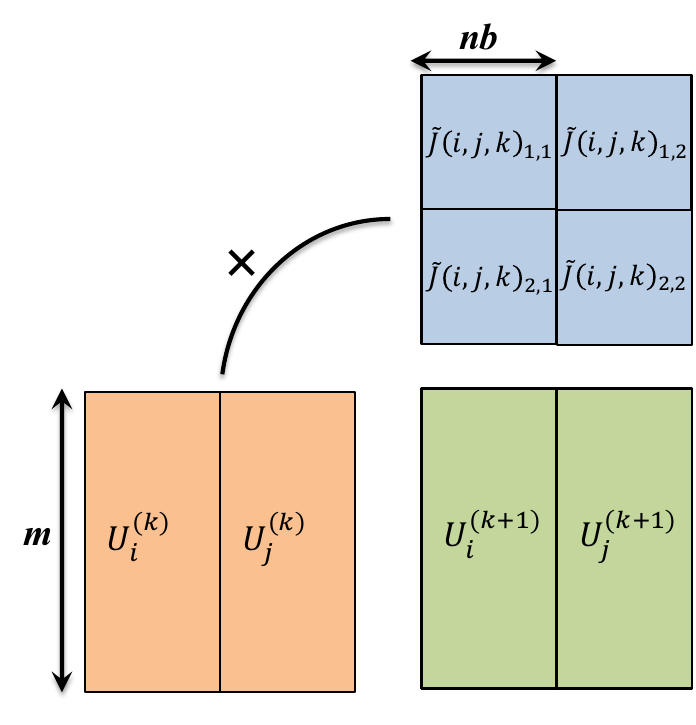}
\caption{GEMM shape for updating the singular vectors.}
\label{fig:gemm_shape_update}
\end{figure}
\subsection{Vector Updates}
\label{subsec:vec}
The singular vector updates can be performed using standard matrix multiplication (\GEMM{}). The eigenvector matrix resulting from the diagonalization of the Gram matrix is used to update both the left and right singular vectors, as shown in \cref{alg:gesvj_unblocked}. \Cref{fig:gemm_shape_update} shows the \GEMM{} shape used in the update. While the eigenvector matrix (i.e., the Jacobi rotation matrix) is contiguous in memory, the block-column pair are not. In order to use the standard batch \GEMM{} routine, we need to perform a two-stage update. As shown in \cref{fig:gemm_shape_update}, these updates are:

\begin{center}
\begin{tabular}{llll}
Stage 1: & $U^{(k+1)}_i \gets U^{(k)}_i \Jb{i}{j}{k}{1,1}$
& \quad & $U^{(k+1)}_j \gets U^{(k)}_i \Jb{i}{j}{k}{1,2}$ \\
Stage 2: & $U^{(k+1)}_i \gets U^{(k+1)}_i + U^{(k)}_j \Jb{i}{j}{k}{2,1}$
& \quad & $U^{(k+1)}_j \gets U^{(k+1)}_i + U^{(k)}_j \Jb{i}{j}{k}{2,2}$ \\
\end{tabular}
\end{center}
The same update sequence applies to the right singular vectors $V$. For each pair $(i,j)$, each stage involves $\bc \cdot \tt{batch}$ matrix multiplications, which can be performed concurrently using batch \GEMM{} with the proper setup of pointer arrays. If $A$ is square, then $U$ and $V$ have the same dimension, and then the updates of $U$ and $V$ can also be combined for increased parallelism, with a total of $2\bc\times \tt{batch}$ \GEMM{} operations per stage. Otherwise, two separate updates for $U$ and $V$ are performed. As in \cref{subsec:gram}, we observe that MAGMA's batch \GEMM{} routine outperforms the vendor's BLAS implementation for small values of $\nb$, and we provide a decision layer to select the best performing routine.
We do not provide a detailed performance comparison between the two routines, as this would be broadly similar to that in \cref{fig:gemm_cn_speedup}.

\subsection{Performance of the Baseline Implementation}
\label{subsec:baseline_perf}
We now present the performance of the baseline implementation. The batch SVD solver is evaluated using square matrices of sizes up to 512, with each dimension tested $10$ times for stable timings. Each matrix is generated randomly using a uniform distribution in the interval $[0,1]$. The accuracy of the decomposition is assessed by checking that the result meets the following accuracy thresholds, where $u$ denotes the unit roundoff and $k$ is a tolerance parameter typically set to 30.
\begin{enumerate}[nolistsep,label=T\arabic*.,ref=T\arabic*]
    \item $\lVert A-USV^H\rVert_1 / (n\lVert A\rVert_1) < ku$. This check is performed using the LAPACK function \tt{BDT01}, executed on the CPU using a LAPACK library such as Intel oneMKL~\cite{mkl} or OpenBLAS~\cite{xqy12,wzzy13,openblas}.\label{it:t1}
    \item $\lVert I-U^H U\rVert_1 / m < ku $ and $\lVert I-V^H V\rVert_1 / n < ku$. This orthogonality check is performed on the CPU using the functions \tt{UNT01} (\tt{ORT01} for real matrices) from the reference implementation of LAPACK.\label{it:t2}
    \item $\lVert \Sigma - \Sigma_{\text{ref}}\rVert_\text{F} / (\min(m,n) \lVert \Sigma_{\text{ref}}\rVert_\text{F})  < ku$, where $\Sigma_{\text{ref}}$ is computed on the CPU with the reference implementation of LAPACK.\label{it:t3}
    \item The $\min(m,n)$ singular values are sorted in descending order.\label{it:t4}
\end{enumerate}

\begin{figure}[!htb]
\centering
\includegraphics[width=0.49\linewidth]{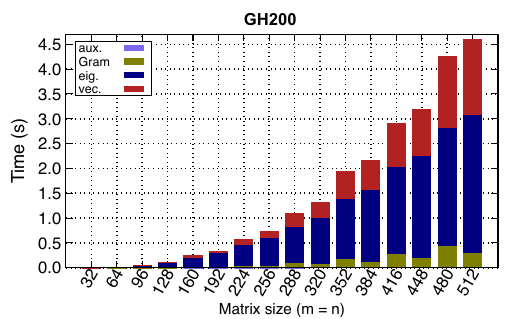}
\includegraphics[width=0.49\linewidth]{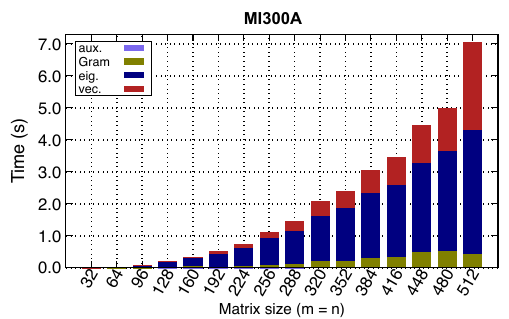}
\includegraphics[width=0.49\linewidth]{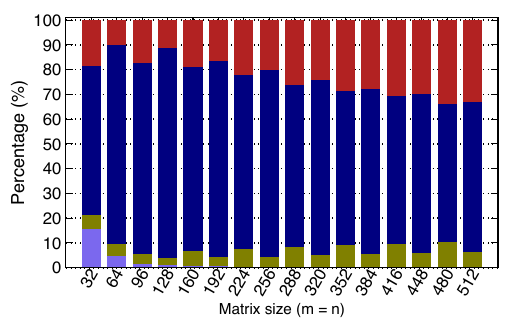}
\includegraphics[width=0.49\linewidth]{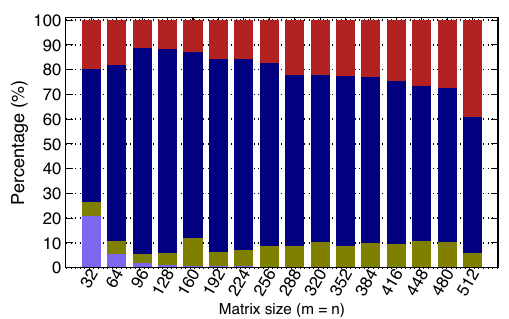}
\caption{Time breakdown of the baseline batch SVD solver on the GH200 system (left) and on the MI300A APU (right). Experiments are conducted in double precision arithmetic.}
\label{fig:baseline_perf}
\end{figure}

\Cref{fig:baseline_perf} shows the timing breakdown of the baseline solver on the GH200 and MI300A systems. All singular values and vectors are computed, and all decompositions satisfy all four accuracy checks~\cref{it:t1}--\ref{it:t4} above. We group the computational steps in four categories.
\begin{enumerate}[nolistsep]
    \item \textbf{aux.}: auxiliary kernels such as setting up the pointer arrays for the batch \GEMM{} kernel, testing for convergence, and finalizing singular values and vectors (normalization, sorting, and vector reordering).
    \item \textbf{Gram}: computation of the Gram matrices, which is performed using the batch \GEMM{} kernel.
    \item \textbf{eig.}: eigendecomposition of the Gram matrices, which is performed using the eigensolver described in \cref{subsec:heevj}.
    \item \textbf{vec.}: vector updates, also performed using the batch \GEMM{} operation.
\end{enumerate}

The Hermitian eigensolver dominates the computation, accounting for 55.8--84.8\% of the total execution time on the GH200 system for and 53.7--83.1\% on the MI300A APU. The vector updates account for over $30\%$ of the time in both cases. The following sections address further optimizations to enhance the time-to-solution.

\section{Inexact Hermitian Eigendecomposition}
\label{sec:inexact_eig}

\subsection{Motivation}
\label{subsec:eig_motivation}
The Hermitian eigensolver dominates the computation in the baseline implementation. In order to reduce the time spent in this step, we could use a faster eigensolver, but we are not aware of a robust batch Hermitian eigensolver that supports all four precisions and is faster than those shown in \cref{fig:heevj_perf}.

Therefore, we explore the possibility of using an inexact (or partial) diagonalization of the Gram matrices. Intuitively, a full orthogonalization of the block columns is not necessary, because orthogonality of the block-column pair is not necessarily preserved by the Jacobi sweep.
To illustrate this further, we show a real numerical example using an $8\times 8$ matrix 
whose entries are shown using four decimal places after rounding. Consider the matrix,

\begin{equation*}
A = 
\begin{bmatrix}
   0.1206 & 0.7675 & 0.3103 & 0.3527 & 0.7382 & 0.7008 & 0.6985 & 0.6836 \\
   0.6438 & 0.8468 & 0.4922 & 0.1086 & 0.8833 & 0.9463 & 0.4762 & 0.0463 \\
   0.0623 & 0.1681 & 0.0378 & 0.8734 & 0.3093 & 0.4652 & 0.1508 & 0.0702 \\
   0.4903 & 0.4045 & 0.6989 & 0.9629 & 0.4463 & 0.3890 & 0.5055 & 0.4994 \\
   0.3061 & 0.3025 & 0.1704 & 0.5332 & 0.0403 & 0.4388 & 0.8133 & 0.2996 \\
   0.8164 & 0.7730 & 0.4167 & 0.4056 & 0.9273 & 0.3014 & 0.1878 & 0.6929 \\
   0.9972 & 0.3156 & 0.1199 & 0.8503 & 0.7538 & 0.8448 & 0.3805 & 0.0510 \\
   0.4246 & 0.8355 & 0.2274 & 0.1604 & 0.5861 & 0.0802 & 0.5890 & 0.6763 \\
\end{bmatrix}.
\end{equation*}
We begin with $\Ak[0] = A$. Using a blocking size $\nb = 2$, our focus will be on the first four columns (or two block columns) of $A$. Using the MATLAB notation for specifying columns of matrices, the first sweep pairs $\Ak[0]_1 = A[,1:2]$ with $\Ak[0]_2 = A[:, 3:4]$ and $\Ak[0]_3 = A[:,5:6]$ with $\Ak[0]_4 = A[:,7:8]$. We then solve two independent Hermitian eigenvalue problems,
\begin{center}
\begin{tabular}{lll}
$G^{(0)}_{1,2}$ & = & $[\Ak[0]_1\;\;\Ak[0]_2]^T[\Ak[0]_1\;\;\Ak[0]_2]$ \\
$G^{(0)}_{3,4}$ & = & $[\Ak[0]_3\;\;\Ak[0]_4]^T[\Ak[0]_3\;\;\Ak[0]_4]$
\end{tabular}
\end{center}

Applying the eigenvectors of $G^{(0)}_{1,2}$ to $[\Ak[0]_1\;\;\Ak[0]_2]$, we obtain
\begin{equation*}
[\Ak[1]_1 \;\; \Ak[1]_2] = 
\begin{bmatrix*}[r]
-0.3726 & 0.7862 & -0.1163 & -0.2318 \\
 0.0407 & 1.0441 &  0.0937 & -0.5339 \\
-0.2475 & 0.6052 & -0.1893 &  0.5771 \\
-0.2084 & 1.2383 &  0.3061 &  0.3863 \\
-0.0477 & 0.6753 & -0.0523 &  0.1935 \\
 0.1593 & 1.2286 &  0.0212 & -0.2586 \\
 0.4584 & 1.2073 & -0.1027 &  0.3913 \\
-0.0684 & 0.8580 & -0.1605 & -0.4349 \\
\end{bmatrix*},\quad
G^{(1)}_{1,2} = 
\begin{bmatrix*}[r]
   0.4876 &      0. &      0. &      0. \\
       0. &  7.7671 &      0. &      0. \\
       0. &      0. &  0.1913 &      0. \\
       0. &      0. &      0. &  1.2676 \\
\end{bmatrix*}
\end{equation*}

So block columns $\Ak[1]_1$ and $\Ak[1]_2$ are now mutually orthogonal, which is expected. However, the next iteration destroys some of this orthogonality. In the next iteration, the pairs $[\Ak[1]_1\;\;\Ak[1]_4]$ and $[\Ak[1]_2\;\;\Ak[1]_3]$ are considered, so that $\Ak[1]_1$ is updated using the eigenvectors of $G^{(1)}_{1,4}$, while $\Ak[1]_2$ is updated with the eigenvectors of $G^{(1)}_{2,3}$. After the vector updates, the orthogonality of $[\Ak[1]_1\;\;\Ak[1]_2]$ is partially lost, since we get
\begin{equation*}
[\Ak[2]_1 \;\; \Ak[2]_2] = 
\begin{bmatrix*}[r]
-0.2854 & 0.7867 & -0.0874 & -0.1007 \\
-0.1634 & 1.0425 &  0.1162 & -0.4349 \\
-0.3304 & 0.6044 & -0.1818 &  0.5698 \\
-0.1142 & 1.2389 &  0.3241 &  0.4416 \\
 0.0112 & 0.6758 & -0.0141 &  0.4473 \\
 0.2488 & 1.2295 &  0.0201 & -0.3562 \\
 0.2259 & 1.2057 & -0.0864 &  0.4151 \\
 0.1746 & 0.8598 & -0.1440 & -0.3579 \\
\end{bmatrix*},\quad
G^{(2)}_{1,2} = 
\begin{bmatrix*}[r]
    0.3739 & -0.0001 & -0.0108 & -0.1912 \\
   -0.0001 &  7.7672 &  0.1312 &  0.4160 \\
   -0.0108 &  0.1312 &  0.1880 &  0.0000 \\
   -0.1912 &  0.4160 &  0.0000 &  1.3463 \\
\end{bmatrix*}.
\end{equation*}

The observation that the orthogonality of a given pair is lost motivates us to consider executing an inexact Hermitian eigensolver. We can achieve this in two ways.
\begin{enumerate}[nolistsep]
    \item We can control the tolerance used inside the Hermitian eigensolver by changing the parameter $k$ in \cref{alg:jacobi_unblocked}. A simple choice is to start with a relatively large $k$ and reduce it at a constant rate until it reaches a value that satisfies the accuracy criteria in \cref{subsec:baseline_perf}.
    \item We can fix the number of sweeps of the Hermitian eigensolver regardless of whether convergence is achieved.
\end{enumerate}

We implemented the dynamic tolerance strategy determining the starting value of $k$ and its reduction rate empirically. Our implementation achieved the required accuracy but yielded a performance gain not exceeding $2\%$. On the other hand, limiting the number of sweeps of the Hermitian eigensolver achieved significant speedup without affecting the overall accuracy of the batch SVD solver. Before showing the performance results, we discuss the numerical behavior of the Jacobi SVD method with an inexact Hermitian eigensolver that uses a fixed number of sweeps.

\subsection{Convergence of the Jacobi SVD Method Using a Single-Sweep Hermitian Eigensolver}
If we use the Jacobi eigensolver to diagonalize the Gram matrix, then running a single sweep of the inner solver instead of running to convergence will not affect the convergence of the SVD solver.
Recall that if every $\Js{i}{j}{k}$ in~\eqref{eq:block-gram} satisfies \ref{it:bVij-1} and \ref{it:bVij-2}, then the one-sided block Jacobi solver is globally convergent.

A single sweep is sufficient to satisfy \ref{it:bVij-2}, as we now explain.
The matrix $\Jl{i}{j}{k}$ is a finite product of Jacobi rotations $J_{i}$, each of which will satisfy
\begin{equation}\label{eq:jacobi-contr}
\off(J^{H} G_{ij} J) < \rho \off(G_{ij}), \qquad 0 \le \rho < 1.
\end{equation}
The per-cycle contraction factor $\rho$ in~\eqref{eq:jacobi-contr} can be arbitrarily close to 1 for some cyclic orderings: for the symmetric eigenvalue problem, for example, Begović Kovač and Hari~\cite{beha17b} show that for $n \ge 4$, one can find a symmetric matrix of order $n$ and a cyclic strategy for which $\rho$ can be arbitrarily close to 1.

For generalized serial strategies, in contrast, Hari and Begović Kovač (2016) establish a uniform per-cycle contraction bound $\rho < 1$ depending only on the matrix dimension nnn, and the broader block-Jacobi generalization of Hari and Begović Kovač (2017) further confirms that serial-type and quasi-serial classes guarantee strict off-norm reduction independently of the input matrix.

Chaining these ensures that
\begin{equation*}
\off(\Js{i}{j}{k}^H G_{i,j} \Js{i}{j}{k}) < \rho \off(G_{i,j}),\qquad 0 \le \rho < 1,
\end{equation*}
with strict inequality whenever $\off(G_{i,j}) > 0$. Hence, \ref{it:bVij-2} holds.

To the best of our knowledge, the only technique that can ensure that the eigenvectors of the Gram matrices will satisfy~\ref{it:bVij-1} was proposed by Drmač~\cite{drma09}, who suggested to compute a Businger--Golub column-pivoted QR factorization of the first block row of $\Js{i}{j}{k}$.
The pivoting will ensure that the singular values of the diagonal blocks of $\Js{i}{j}{k}$ are bounded below by a constant depending only on the block size.
However, this factorization is extremely onerous in terms of data movement, and we are not aware of any block Jacobi eigensolver that performs this step to ensure convergence.
Therefore, we have decided to skip this step in our implementation. Later in the paper, we show that, in practice, the proposed solver converges on a wide range of problems.

\subsection{Impact of the Inexact Eigensolver on Performance}
\label{subsec:perf_1sweep}
Since reducing the number of sweeps of the Hermitian eigensolver does not affect the convergence of the one-sided Jacobi SVD method, we now study the performance of the algorithm when only one sweep of the eigensolver is performed.
The updated batch SVD solver, called \emph{Design-2}, satisfies the four accuracy thresholds \ref{it:t1}--\ref{it:t4} in \cref{subsec:baseline_perf}---the numerical accuracy of the solver will be discussed in more detail in \cref{sec:final_perf}.
\begin{figure}[!hb]
\centering
\includegraphics[width=0.49\linewidth]{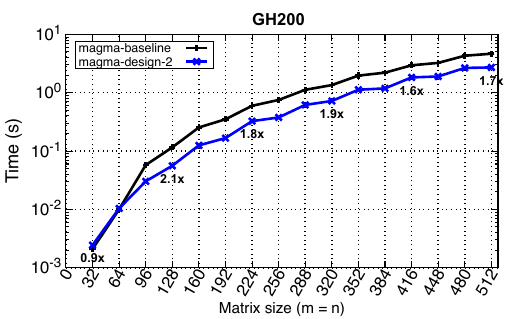}
\includegraphics[width=0.49\linewidth]{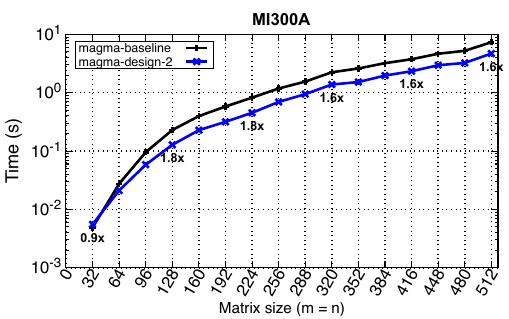}
\caption{Performance gain of Design-2 over the baseline of the batch SVD solver on the GH200 system (left) and on the MI300A APU (right). Experiments are conducted in double precision arithmetic using a batch of 1000 square matrices.}
\label{fig:speedup_sweep}
\end{figure}

\Cref{fig:speedup_sweep} shows the performance of Design-2 over the MAGMA baseline on square matrices of size up to $512$. The new solver is between 1.6 and 2.2 times faster than the baseline on both GPUs, while achieving the same accuracy.
For very small matrices, we observe a slowdown, because the baseline can compute the SVD with only one call to the Hermitian eigenvsolver. In this case, the computed eigenvectors correspond to the right singular vectors $V$, and the identity $U\Sigma = AV$ can be used to compute the singular values and left singular vectors. Using an inexact eigensolver on such small matrices leads to multiple Jacobi SVD sweeps and to a larger time-to-solution overall.
\begin{figure}[!htb]
\centering
\includegraphics[width=0.49\linewidth]{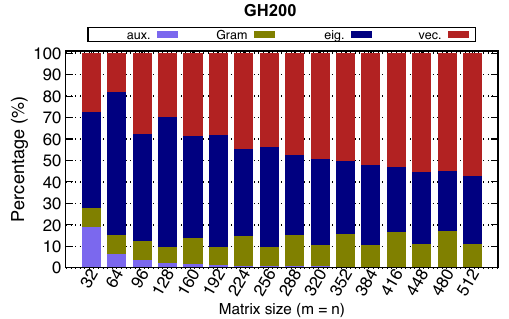}
\includegraphics[width=0.49\linewidth]{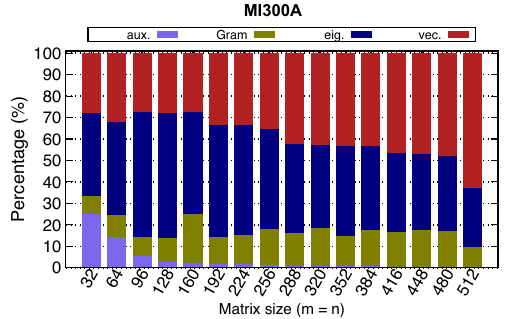}
\caption{Time breakdown of Design-2 of the batch SVD solver on the GH200 system (left) and on the MI300A APU (right). Experiments are conducted in double precision arithmetic.}
\label{fig:design2_breakdown}
\end{figure}

\cref{fig:design2_breakdown} shows a time breakdown for Design-2. The impact of the Hermitian eigensolver is significantly reduced: on the GH200 system, the single-sweep eigensolver accounts for 28.1--66.4\% in Design-2, down from 55.8--84.8\% for the baseline; on the MI300A APU, it accounts for 27.5--58.2\% of the total execution time, compared with 53.7--83.1\% for the baseline. The vector updates now dominate the computation, accounting for over 50\% of the total on both GPUs for the largest matrices.
In the next section, we discuss how the design can be optimized for the vector updates.

\section{Optimized Vector Updates}
\label{sec:vec_updates}
The baseline design uses standard batch \tt{GEMM} operations, as described in \cref{subsec:vec}. A shortcoming of this approach is that the batch \tt{GEMM} kernels are called on the skewed shapes shown in \cref{fig:gemm_shape_update}, for which they may not be properly optimized. In addition, the vector updates are performed in two calls to batch \GEMM, leading to unnecessary memory traffic. We now explain how a customized implementation can reduce memory traffic and thereby improve the performance of our solver.

For a pair of columns $U_{i}$ and $U_{j}$, the update can be performed in two stages.
\begin{center}
\begin{tabular}{llll}
Stage 1: & $U^{(k+1)}_i \gets U^{(k)}_i \Jb{i}{j}{k}{1,1}$ & \quad & $U^{(k+1)}_j \gets U^{(k)}_i \Jb{i}{j}{k}{1,2}$ \\
Stage 2: & $U^{(k+1)}_i \gets U^{(k+1)}_i + U^{(k)}_j \Jb{i}{j}{k}{2,1}$ & \quad & $U^{(k+1)}_j \gets U^{(k+1)}_j + U^{(k)}_j \Jb{i}{j}{k}{2,2}$ \\
\end{tabular}
\end{center}
The two multiplications in each stage can be performed concurrently, but a standard \tt{GEMM} implementation will create unnecessary memory traffic between the two stages. On the one hand, one must load the two original block columns $U^{(k)}_i$ and $U^{(k)}_j$ twice, as each is used once for $U^{(k+1)}_i$ and once for $U^{(k+1)}_j$. On the other hand, one must store the intermediate results of $U^{(k+1)}_i$ and $U^{(k+1)}_j$ in the global memory at the end of Stage 1 and load them again in Stage 2.

\begin{algorithm}[!htb]
\KwData{$U^{(k)}_{i} \in \Cmnb$, $U^{(k)}_{j} \in \Cmnb$, $\Js{i}{j}{k} \in \Cnbnb$.}
\KwParam{Blocking size $mb$.}
\KwResult{Updated vectors $U^{(k+1)}_{i} \in \Cmnb$, $U^{(k+1)}_{j} \in \Cmnb$.}
\tcp{Initialize register buffers, each of size $mb\times nb$.}
$rU_{i} \gets 0$\;
$rU_{j} \gets 0$\;
\BlankLine
\tcp{Point to corresponding block in global memory.}
$pU_{i} \gets U^{(k)}_i + \tt{blockId}\times mb$ \;
$pU_{j} \gets U^{(k)}_j + \tt{blockId}\times mb$ \;
\BlankLine
\tcp{Load data in shared memory.}
$sU \gets \tt{load(}pU_{i}\tt{)}$
\tcp*{\parbox{8.5cm}{$mb \times \nb$}}
$sJ \gets \tt{load(}[\Jb{i}{j}{k}{1,1}\quad \Jb{i}{j}{k}{1,2}]\tt{)}$
\tcp*{\parbox{8.5cm}{$\nb \times 2\nb$}}
\BlankLine
\tcp{First multiplication}
$[rU_i\quad rU_j] \pluseq sU \times sJ$ \;
\BlankLine
\tcp{Load data in shared memory.}
$sU \gets \tt{load(}pU_{j}\tt{)}$
\tcp*{\parbox{8.5cm}{$mb \times \nb$}}
$sJ \gets \tt{load(}[\Jb{i}{j}{k}{2,1}\quad \Jb{i}{j}{k}{2,2}]\tt{)}$
\tcp*{\parbox{8.5cm}{$\nb \times 2\nb$}}
\BlankLine
\tcp{Second multiplication.}
$[rU_i\quad rU_j] \pluseq sU \times sJ$ \;
\BlankLine
\tcp{Overwrite $U^{(k)}_i$ and $U^{(k)}_j$ with $U^{(k+1)}_i$ and $U^{(k+1)}_j$}
$pU_i \gets \tt{store(}rU_{i}\tt{)}$\;
$pU_j \gets \tt{store(}rU_{j}\tt{)}$\;
\caption{One thread-block in the Vector Update Kernel.}
\label{alg:custom_vec_update}
\end{algorithm}

We address these performance issues with a customized \GEMM-like kernel for updating the singular vectors. The kernel uses the same building blocks (i.e., CUDA/HIP device functions) as MAGMA's batch \GEMM{} kernel, such as moving data blocks between shared memory and global memory or multiplying two data blocks in shared memory. The optimized update kernel reads each block column exactly once. It also merges stages 1 and 2 into one kernel, so that the intermediate results are kept in the register file without being written to global memory. The pseudo code of the update kernel is shown in \cref{alg:custom_vec_update}. Each block-column is partitioned horizontally into $ \left \lceil \frac{m}{\mb} \right \rceil$ blocks, where $\mb$ is a tuning parameter of the kernel. The kernel is launched using a 2D grid of size $\left \lceil \frac{m}{\mb} \right \rceil \times \frac{1}{2}\bc \cdot \tt{batch}$ thread-blocks, where each thread block is assigned a partition of size $\mb\times\nb$ in a given block-column pair.

\begin{figure}[!tb]
\centering
\includegraphics[width=0.49\linewidth]{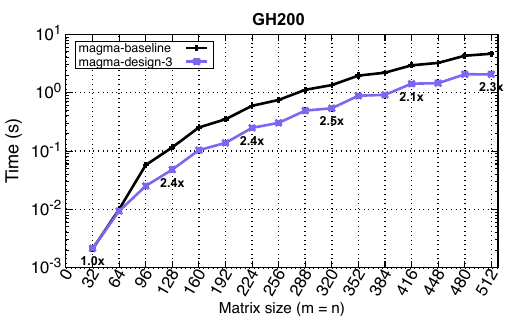}
\includegraphics[width=0.49\linewidth]{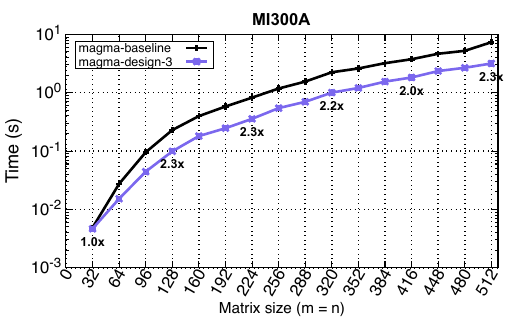}
\caption{Performance gain of Design-3 over the baseline of the batch SVD solver on the GH200 system (left) and on the MI300A APU (right). Experiments are conducted in double precision arithmetic with 1,000 square matrices per batch.}
\label{fig:speedup_vec}
\end{figure}

\emph{Design-3} combines Design-2 with the customized \GEMM{} kernel for the vector updates. Design-3 meets all the accuracy requirements mentioned in \cref{subsec:baseline_perf}, as it does not alter the numerical behavior of Design-2. \Cref{fig:speedup_vec} shows the performance gain of Design-3 over the baseline implementation, which is up to $2.5\times$ on the GH200 system, and up to $2.3\times$ on the MI300A APU. When compared with to Design-2, the optimized vector updates bring speedups of 11--33\% on the GH200 GPU, and of 19--48\% on the MI300A APU.


\section{Masked Batch Operations}
\label{sec:batch_mask}
Our final optimization focuses on the varying convergence speed of the problems within the same batch. The SVD solver is an iterative algorithm, and the number of steps needed to converge can vary significantly, even for matrices of the same dimensions. This means that matrices with early convergence could undergo unnecessary computation.

As an example, consider a batch of two matrices, $A_e$ and $A_l$. If $A_e$ requires fewer Jacobi sweeps than $A_l$ to be decomposed, then $A_e$ converges earlier than $A_l$. 
The batch solver we have described thus far only detects global convergence at the batch level, thus Jacobi sweeps keep being applied to $A_e$ until $A_l$ has converged. During these extra sweeps, the Gram matrices computed for $A_e$ are all diagonal. The Hermitian eigensolver is applied to these diagonal matrices, producing identity eigenvectors, which are applied during the vector updates. This additional computation does not affect the accuracy of the SVD of $A_e$, but it may incur a substantial performance and energy overhead.

In principle, one could implement a mechanism that detects converged problems, removes them from the batch, and continues with a new smaller batch. Such a mechanism, however, could introduce an overhead that yields a slowdown rather than a performance gain. Therefore, we adopt an alternative strategy based on \emph{masking} within batch operations. Specifically, we use \emph{batch masks} to select which problems in the batch can terminate early. In the example above, we could mask the workload of $A_e$ off the batch as soon as the computation of its SVD has converged, saving the unnecessary compute. This approach is lightweight, broadly applicable, and has the potential to become a standard feature in batch linear algebra.

We implement masking for the in-house batch kernels we developed for this work, namely the Hermitian eigensolver and the optimized vector updates.
Existing batch BLAS interfaces do not support masks, thus we could not apply this optimization to the computation of the Gram matrix, which partially relies on the vendor library.

In the Hermitian eigensolver, we mask off a problem when all block-column pairs in a single sweep are found to be orthogonal. This could be detected by inspecting all the Gram matrices prior to calling the eigensolver, but it would incur a significant read overhead. We opt to check the output information of the eigensolver from the previous Jacobi sweep. In particular, we use (1) a flag indicating whether the eigensolver has converged, and (2) the number of sweeps executed. If these indicate that convergence was reached after exactly one sweep, then we conclude that the input matrix was already diagonal and the corresponding columns already orthogonal. This information can help mask off the corresponding eigenvalue problems in subsequent sweeps, as well as the corresponding vector updates, since the Jacobi rotation is nothing more than the identity matrix.

\begin{figure}[!htb]
\centering
\includegraphics[width=0.49\linewidth]{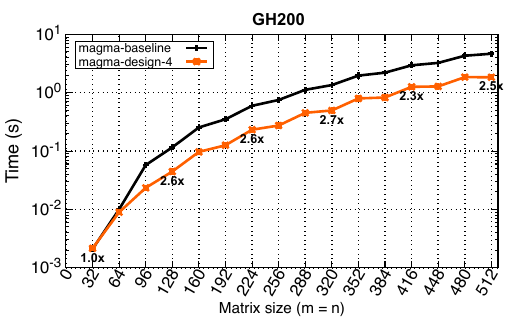}
\includegraphics[width=0.49\linewidth]{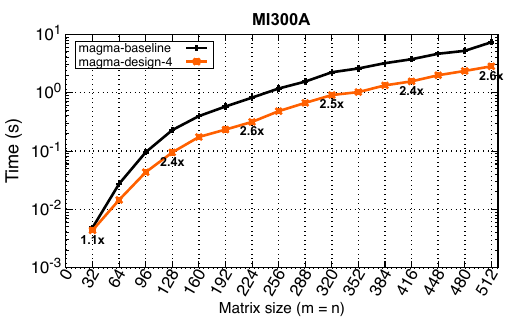}
\caption{Performance gain of Design-4 over the MAGMA baseline on the GH200 system (left) and on the MI300A APU (right). Experiments are conducted in double precision arithmetic with 1,000 square matrices per batch.}
\label{fig:speedup_mask}
\end{figure}

\emph{Design-4} is the batch SVD solver that adds the masked batch operations to Design-3. Design-4 does not alter the numerical behavior of the solver and meets the accuracy thresholds in \Cref{subsec:baseline_perf}. \Cref{fig:speedup_mask} shows that Design-4 is up to 2.7 times faster than the baseline on the GH200 system, and up to 2.6 times faster on the MI300A APU. When compared with Design-3, Design-4 achieves up to 14\% speedup on the GH200 system and up to 18\% speedup on the MI300A APU.

\section{Other Performance Considerations}
\label{sec:special}
If the matrices in the batch are very small or have an unusually skewed shape, alternative execution paths may yield better performance.

\subsection{Optimizations for very small problems}
\label{subsec:fused}
If the problem is small enough to fit in the shared memory of the GPU along with any required workspace, then it is common practice to launch a single kernel for the entire operation~\cite{ahtd17,abdelfattah2022batch}. 
This alternative execution path ensures an optimal memory traffic, because the input problem is read exactly once from the global memory, and the output values and vectors are also written only once. We implement an unblocked version of the one-sided Jacobi SVD solver in \cref{alg:gesvj_unblocked}, which runs entirely in the shared memory of the GPU. The kernel supports square and rectangular shapes, and is selected for execution at runtime if the underlying GPU satisfies the shared memory requirements. The performance of this kernel is superior to that of the blocked implementation, as we show in \cref{sec:final_perf}

\subsection{Pre-processing the SVD using a QR Factorization}
\label{subsec:qr}
Consider an input matrix $A\in \Cmn$, and assume that $m > n$.
This assumption is not restrictive, as one can consider $A^H = V \Sigma U^H$ if $ m < n$.  
If $m \gg n$, then a QR factorization of $A$ can be an effective pre-processing step.
Consider the Businger--Golub QR factorization $AP = Q[R^H\;\;0]^H$, where $P \in \Cnn$ is a permutation matrix that improves the numerical stability through column pivoting, $Q \in \Cmn$ is unitary, and $R \in \Cnn$ is upper triangular. This factorization is a rank-revealing QR decomposition where $P$ reorders the columns of $A$ so that the diagonal entries of $R$ decay in magnitude.
Drmač and Veselić have shown that applying the one-sided SVD Jacobi algorithm to the triangular factor $R^H$ can speed up convergence~\cite{drve08I} and reduce both FLOP count and data movement~\cite{drve08II}.

    


Our implementation first computes the non-pivoted factorization $A = Q[R^H\;\;0]^H$, then the SVD $R = \wh{U}\Sigma V^H$, and finally recovers $U = Q\wh{U}$. The performance gain depends on the ratio $m/n$. If $m\gg n$, then it is more efficient to compute the SVD of the small square matrix $R$ rather than the original matrix. This behavior assumes that the QR factorization can be computed efficiently on the GPU, which is the case for the MAGMA library~\cite{abdelfattah2022batch}. A column-pivoted QR factorization would also improve the convergence of the SVD solver, but we are not aware of any efficient GPU implementation. We show the impact of the QR factorization in \cref{sec:final_perf}.
\section{Final Performance Results}
\label{sec:final_perf}

\subsection{Numerical Accuracy}
\label{subsec:accuracy}
We assess the accuracy of the batch SVD solver (from now on, the MAGMA batch SVD solver) by computing four error checks for every test case. These errors, which correspond to the three thresholds \ref{it:t1}--\ref{it:t3} in \cref{subsec:baseline_perf}, are:
\begin{enumerate}[nolistsep,label=$\mathrm{e}_{\arabic*}$., ref=$\mathrm{e}_{\arabic*}$]
    \item $\lVert A-USV^H\rVert_1 / (n\lVert A\rVert_1)$,\label{it:e1}
    \item $\lVert I-U^H U\rVert_1 / m $,\label{it:e2}
    \item $\lVert I-V^H V\rVert_1 / n $,\label{it:e3}
    \item $\lVert \Sigma - \Sigma_{\text{ref}}\rVert_\text{F} / (\min(m,n)\lVert\Sigma_{\text{ref}}\rVert_\text{F})$.\label{it:e4}
\end{enumerate}

The numerical accuracy is checked across all four standard LAPACK data types (single, double, single-complex, and double-complex). For each of the four error measures, we report the maximum error observed across the batch for a given dimension. To ensure the robustness of the solver, we consider the six families of matrices in \cref{tbl:svd_dist}, which present varying distributions of singular values.
If the matrix is not randomly generated, the condition number $\kappa_2(A)$ is set to $10^5$ for single and single-complex matrices, and to $10^{10}$ for double and double-complex matrices.
\begingroup
\begin{table}[t]
\renewcommand{\arraystretch}{1.2} 
\centering
\caption{Matrices used in the accuracy evaluation of the batch SVD solver}
\label{tbl:svd_dist}
\begin{tabular}{lll}
\toprule
\bld{Matrix name} & \bld{Singular value distribution} & \bld{Singular values} ($i = 1, 2, \ldots, n$)\\
\midrule
\tt{random} & Data dependent & \\ 
\tt{arith} & Arithmetic       & $\sigma_i = 1 - \left(\frac{i-1}{n-1}\right) \bigl(1 - \frac{1}{\condt{A}}\bigr)$ \\
\tt{cluster0} & Clustered        & $\sigma_1 = 1$, $\sigma_i = \frac{1}{\condt{A}}$ for $i > 1$ \\
\tt{cluster1} & Clustered        & $\sigma_i = 1$ for $i < n$, $\sigma_n = \frac{1}{\condt{A}}$ \\
\tt{logrand} & Logarithmic      & $\log(\sigma_i)$ uniform on $\left[\log\bigl(\frac{1}{\condt{A}}\bigr), \log(1)\right]$ \\
\tt{geo} & Geometric        & $\sigma_i = \condt{A}^{(\frac{1-i}{n-1})}$ \\
\bottomrule
\end{tabular}
\end{table}
\endgroup

\begin{figure}[!htb]
\centering
\includegraphics[width=0.49\linewidth]{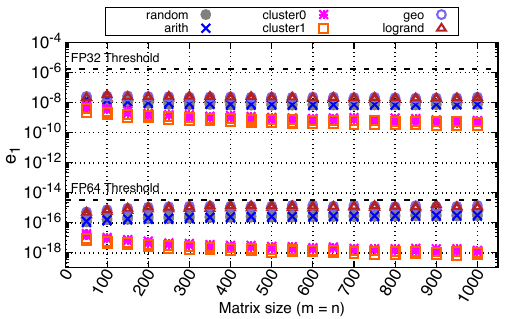}
\includegraphics[width=0.49\linewidth]{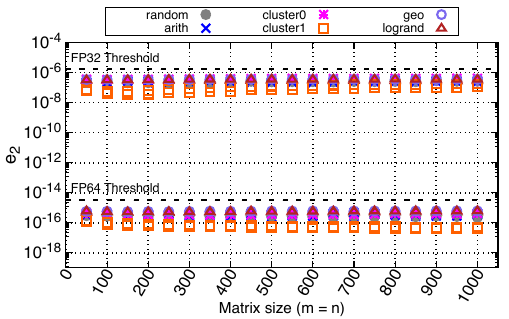}
\includegraphics[width=0.49\linewidth]{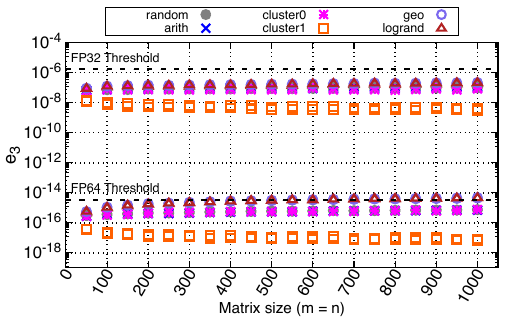}
\includegraphics[width=0.49\linewidth]{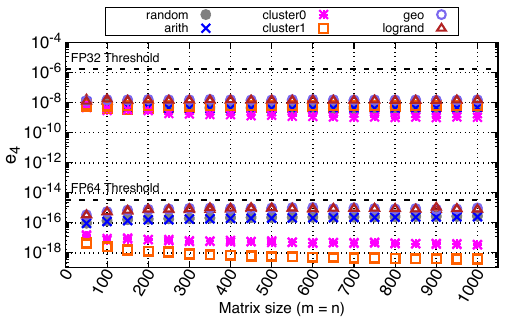}
\caption{Numerical accuracy of the batch SVD solver using the matrix types described in \cref{tbl:svd_dist}. Results are shown for batches of 100 matrices. The top half of each panel shows the accuracy for single(-complex) precision matrices, and the bottom half shows the accuracy for the double(-complex) precision matrices. All experiments are conducted on the GH200 system.}
\label{fig:accuracy}
\end{figure}

\Cref{fig:accuracy} shows the value of \ref{it:e1}, \ref{it:e2}, \ref{it:e3}, and \ref{it:e4} on the GH200 system. The numerical behavior on the AMD APU is similar and therefore not shown. The top half of each panel shows the accuracy for single and single-complex matrices, and the bottom half shows the accuracy for double and double-complex matrices. 
The two values are marked as ``FP32 Threshold'' and ``FP64 Threshold'' are set to $1.7881\times 10^{-6}$ and $3.3307\times 10^{-15}$, respectively, which are equivalent to $ku$, where $k = 30$ and $u$ is the unit roundoff for both FP32 and FP64.
The proposed batch SVD solver satisfies these accuracy thresholds in the large majority of cases. The only exception is the orthogonality $V$, which for double precision slightly exceeds it for the \verb+logrand+ and \verb+geo+ matrices.

\subsection{Comparison with Existing Batch SVD Solvers}
\label{subsec:competition}
Next, we compare the MAGMA batch SVD solver 
against four existing GPU alternatives. 
\begin{enumerate}[nolistsep]
    \item cuSOLVER~\cite{cusolver} is NVIDIA's GPU accelerated numerical library for computing matrix decompositions and solving linear systems. Its batch SVD solver is \verb+cusolverDn<datatype>gesvdj+, where \verb+<datatype>+ can be either `\verb+s+', `\verb+d+', `\verb+c+', or `\verb+z+', following the LAPACK conventional type names. The solver is limited to small matrices of dimensions up to $32$. The character `\verb+j+' at the end of the routine name indicates a Jacobi-based method.
    \item rocSOLVER~\cite{rocsolver} is AMD's implementation of LAPACK routines for dense linear algebra. Its batch SVD solver is \verb+rocsolver_<datatype>gesvd_batched+, where \verb+<datatype>+ can be either `\verb+s+', `\verb+d+', `\verb+c+', or `\verb+z+' as above. This routine first reduces the input matrix to a bi-diagonal form and then uses the QR algorithm.
    \item KBLAS~\cite{akl16,clk16,almk17} is an open-source library for GPUs that implements a subset of BLAS and LAPACK. Its batch SVD solver, \verb+kblas_gesvj_batch+, is an overloaded interface that only supports the single and double data types~\cite{btlk18}. KBLAS uses the one-sided Jacobi SVD algorithm, but instead of computing the Hermitian eigen-decomposition of the Gram matrix, it computes the SVD of a given block-column pair. It also provides an unblocked parallel-Jacobi implementation for very small matrices and a blocked serial-Jacobi implementation for relatively larger matrices. There is no option to compute the right singular vectors, which can be obtained using a batch linear solver. KBLAS does not support AMD GPUs.
    \item Wcycle-SVD~\cite{xpxs22} is an open-source batch SVD solver that uses the one-sided Jacobi algorithm which does not sort the singular values. Algorithmically, the implementation uses a parallel blocked Jacobi algorithm with Hermitian eigen-decomposition of the Gram matrices, which is similar to the solver we developed. Wcycle-SVD provides two interfaces for double-precision matrices only: \verb+svd_small_matrix+ and \verb+svd_large_matrix+. The former interface failed most of the accuracy checks, so we do not include it in our benchmarks, while the left singular values produced by the large-matrix interface occasionally failed the orthogonality test. We were unable to compile Wcycle-SVD on the AMD system.
\end{enumerate}
\begin{figure}[!htb]
\centering
\includegraphics[width=0.49\linewidth]{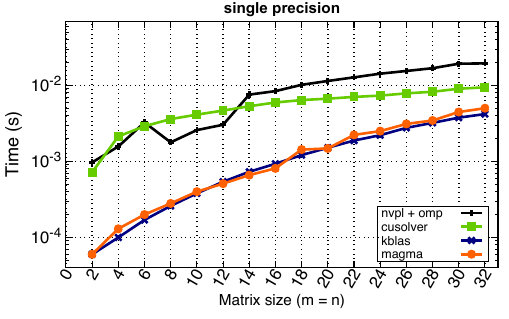}
\includegraphics[width=0.49\linewidth]{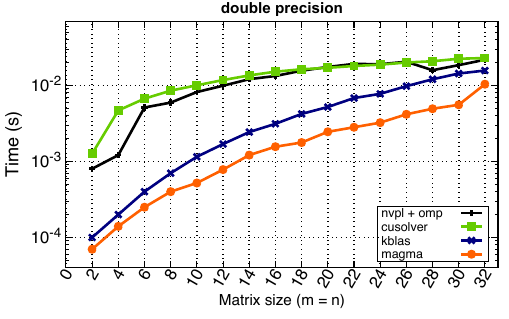}
\includegraphics[width=0.49\linewidth]{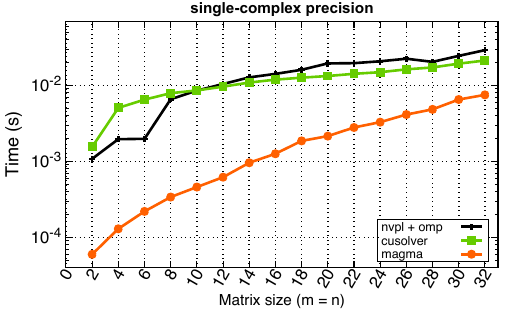}
\includegraphics[width=0.49\linewidth]{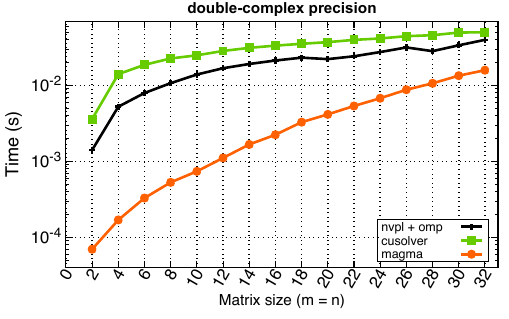}
\caption{Performance of the batch SVD on the GH200 GPU. Results are shown for batches of 10,000 very small matrices.}
\label{fig:final_smallsq_gh200}
\end{figure}

In addition to these four GPU solvers, we also include the performance of two existing parallel LAPACK implementation running on CPU: the NVIDIA Performance Libraries (NVPL) on the GH200 systems, and the Intel Math Kernel Library (MKL) on the MI300A system. On both systems, we benchmark the \verb+gesvd+ routine inside an OpenMP \verb+parallel for+ loop. The number of OpenMP threads is set to 72 on the GH200 system and to 24 on the MI300A system.

To prove the robustness of the proposed SVD solvers, we show comprehensive performance benchmarks using all standard LAPACK precisions. The performance is shown for both square and tall-skinny matrices. For tall-skinny matrices, we show two performance graphs for MAGMA, one with and one without the QR factorization pre-processing. The performance of the two approaches is almost the same on square sizes, and so only one approach is shown (without QR factorization). All matrices are randomly generated using the LAPACK \tt{larnv} routine configured with a uniform distribution on the interval $[0,1]$. 
\subsection{Performance on Very Small Matrices}
\label{subsec:smallsq}
\begin{figure}[!tb]
\centering
\includegraphics[width=0.49\linewidth]{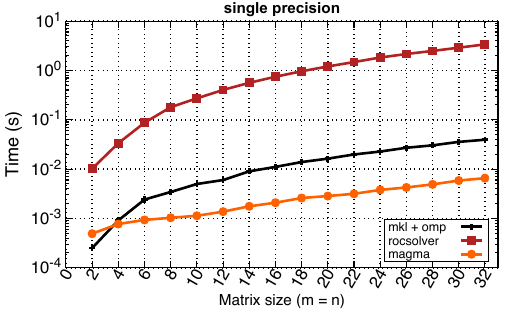}
\includegraphics[width=0.49\linewidth]{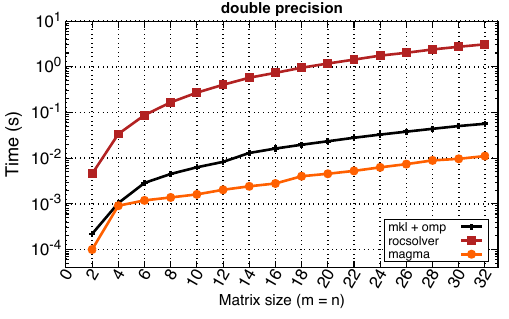}
\includegraphics[width=0.49\linewidth]{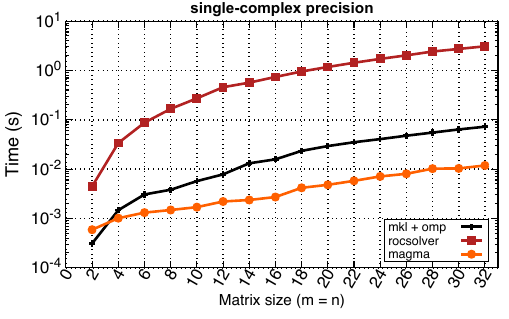}
\includegraphics[width=0.49\linewidth]{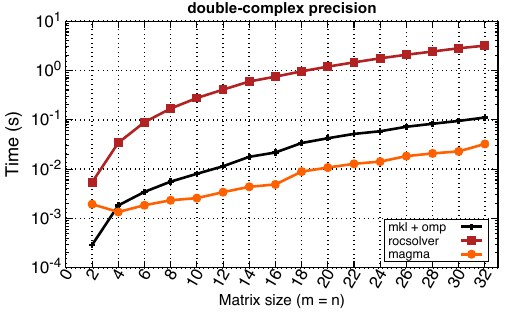}
\caption{Performance of the batch SVD on the MI300A APU. Results are shown for batches of 10,000 very small matrices.}
\label{fig:final_smallsq_mi300a}
\end{figure}

\Cref{fig:final_smallsq_gh200,fig:final_smallsq_mi300a} show the time-to-solution for batches of 10,000 very small matrices on the GH200 and the MI300A systems, respectively. The matrices are small enough that the MAGMA SVD solver defaults to the single-kernel approach in \cref{subsec:fused}. Profiling the execution of the other four GPU solvers suggests that all but rocSOLVER take a similar approach. 

On the GH200 system, the MAGMA SVD solver achieves the best time-to-solution for all data types except single precision, where KBLAS is comparable and often slightly faster than MAGMA. In single precision, MAGMA is 3.9--16.4$\times$ faster than the CPU implementation, and 1.9--16.9$\times$ faster than cuSOLVER. In double precision, the MAGMA SVD solver is 2.1--20.5$\times$ faster than the CPU implementation, 2.2--33.4$\times$ faster than cuSOLVER, and 1.4--2.6$\times$ faster than KBLAS. In single-complex precision, the MAGMA SVD solver is 3.7--19.0$\times$ faster than the CPU implementation and 2.8--39.0$\times$ faster cuSOLVER. A similar performance trend is also observed in the double-complex precision case, with a 2.5--30.0$\times$ speedup over the CPU implementation and a 3.2--80.6$\times$ speedup over cuSOLVER. Wcycle-SVD is not considered as it failed to execute for such small matrices. The performance results of KBLAS do not include the computation of the right singular vectors. 

On the MI300A system, the CPU implementation is generally faster than rocSOLVER. As we mentioned before, rocSOLVER applies its general (i.e., non-fused) solver even for very small matrices, which leads to additional memory traffic between different computational stages. As a result, MAGMA is over two orders of magnitude faster than rocSOLVER and up to $6.4\times$ faster than the CPU implementation across all LAPACK data types. 

\subsection{Performance on Relatively Large Square Matrices}
\label{subsec:bigsq}
\begin{figure}[!htb]
\centering
\includegraphics[width=0.49\linewidth]{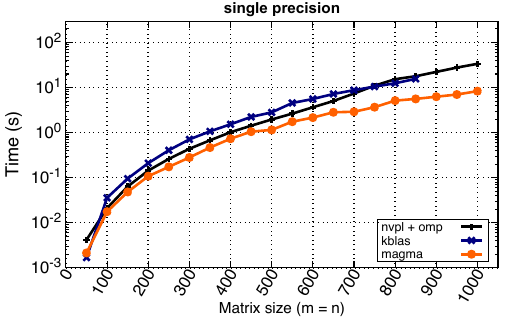}
\includegraphics[width=0.49\linewidth]{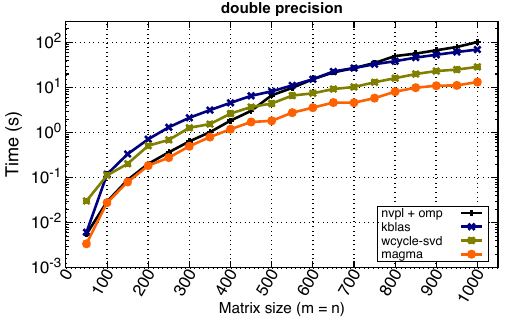}\\
\vspace{1em}
\includegraphics[width=0.49\linewidth]{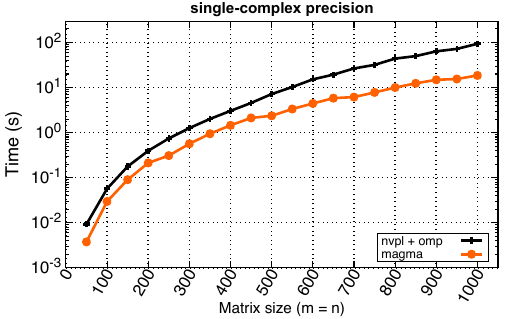}
\includegraphics[width=0.49\linewidth]{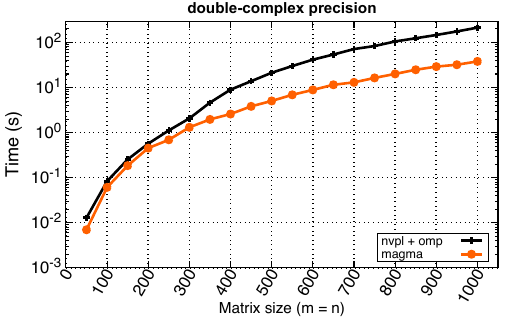}
\caption{Performance of the batch SVD on the GH200 system. Results are shown for batches of 1,000 square matrices.}
\label{fig:final_sq_gh200}
\end{figure}

We extend the benchmark on square matrices to accommodate larger matrices of order up to 1,000. In order to avoid very long runtimes, we use smaller batches of $1,000$ matrices each. \Cref{fig:final_sq_gh200} shows the performance on the GH200 system. The MAGMA SVD solver is the best performing GPU solution across all precisions. The speedup against the CPU is up to $7.7\times$. MAGMA outperforms KBLAS by up to $3\times$ in single precision and up to 5.9$\times$ in double precision. MAGMA is also 1.9--8.9$\times$ faster than WcycleSVD. Recall that cuSOLVER does not support this range of sizes. In addition, KBLAS encountered a run-time error for single precision on sizes larger than 850.

\begin{figure}[!htb]
\centering
\includegraphics[width=0.49\linewidth]{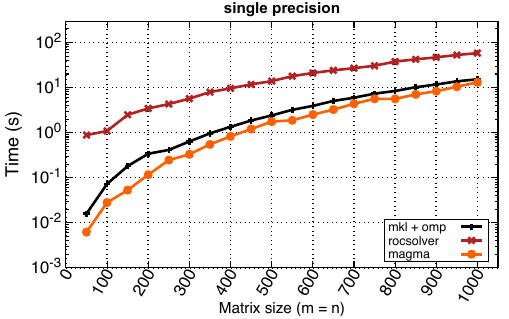}
\includegraphics[width=0.49\linewidth]{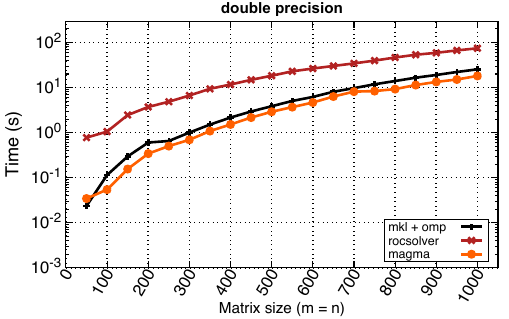}\\
\vspace{1em}
\includegraphics[width=0.49\linewidth]{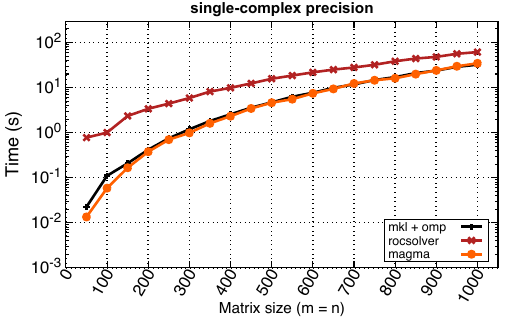}
\includegraphics[width=0.49\linewidth]{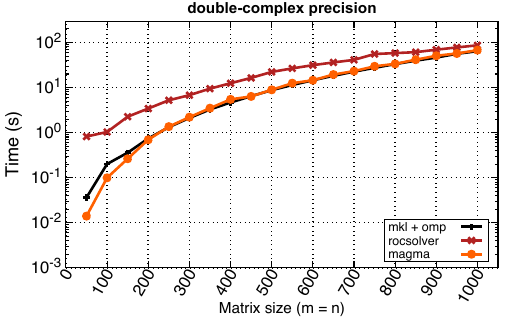}
\caption{Performance of the batch SVD on the MI300A APU. Results are shown for batches of 1,000 square matrices.}
\label{fig:final_sq_mi300a}
\end{figure}

\Cref{fig:final_sq_mi300a} shows the performance on the MI300A APU. MAGMA still achieves the best
time-to-solution across all data types, with the MKL solver having a very similar performance.The MAGMA SVD solver is significantly faster than rocSOLVER, achieving speedups up to $142.5\times$, $22.5\times$, $57.7\times$, and $58.4\times$ for single, single-complex, double, and double-complex data types, respectively. 

We observe performance gap between MAGMA and rocSOLVER becomes asymptotically smaller for all data types. In fact, the minimum speedup against rocSOLVER is observed at the largest dimensions, with a minimum speedup of $4.5\times$ for single, $4.1\times$ for double, $1.8\times$ for single-complex, and $1.3\times$ for double-complex. This is due to two important factors.

First, increasing the precision increases the register pressure and shared memory requirements for different computational kernels. This effect limits the overall occupancy on the GPU in terms of the number of concurrent computations residing on the same multiprocessor/compute unit. Second, the Hermitian eigensolver needs to cache two matrices of size $2nb\times 2nb$ in shared memory. Therefore, increasing the precision will limit the value of $nb$, which will in turn increase the number of block-column pairs, and consequently the number of iterations in a Jacobi sweep.

These two observations apply to both systems, but they will have a greater impact on performance on the MI300A APU, which has a significantly smaller shared memory---64KB, compared with the 227KB of the GH200 system. They explain the drop in asymptotic speedups between single and double precision and between single-complex and double-complex precision in \cref{fig:final_sq_mi300a}. The double and single-complex data types have similar memory requirements, and in this case the drop in performance is due to the fact that the computation of the Gram matrix is considerably more time-consuming for complex data types compared with real ones, which leads to a slowdown in the overall time-to-solution. 

\subsection{Performance on Tall-Skinny Matrices}
\label{subsec:tallskinny}
In this benchmark, we fix the number of columns to 16, and vary the number of rows between 100 to 2,000. This is a compelling case to show the impact of the QR factorization as a pre-processing stage. The KBLAS and rocSOLVER libraries apply this strategy by default regardless of the shape of the matrix. MAGMA offers two separate interfaces for the batch SVD solver, so that the end user has full control over the numerical behavior of the solver and can decide whether the QR factorization is necessary.   
\begin{figure}[!htb]
\centering
\includegraphics[width=0.49\linewidth]{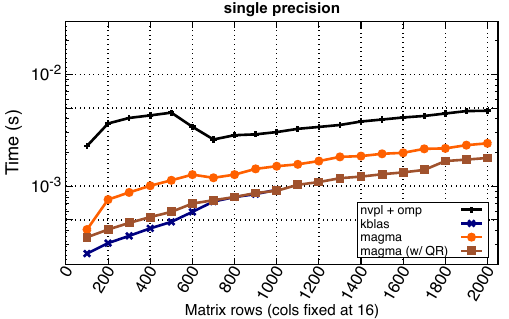}
\includegraphics[width=0.49\linewidth]{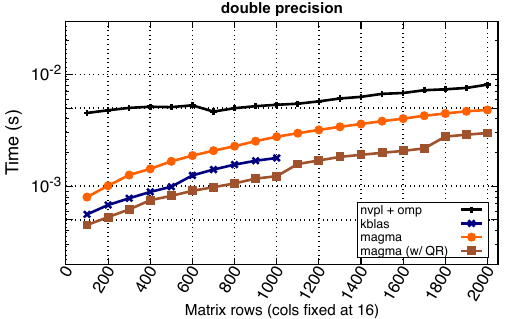}\\
\vspace{1em}
\includegraphics[width=0.49\linewidth]{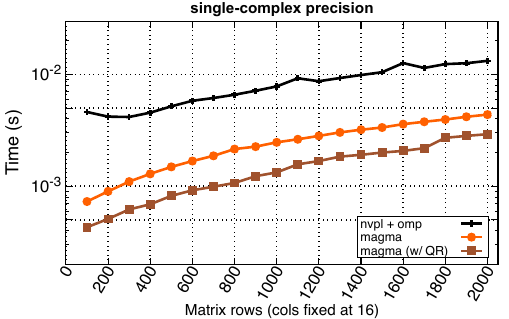}
\includegraphics[width=0.49\linewidth]{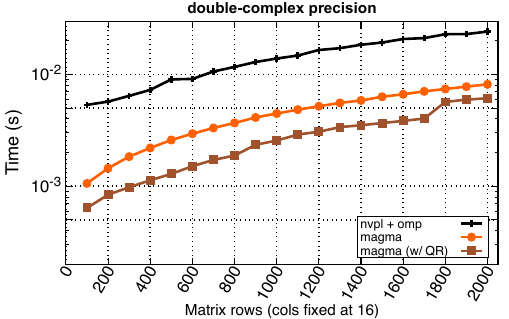}
\caption{Performance of the batch SVD on the GH200 system. Results are shown for batches of 1,000 tall-skinny matrices.}
\label{fig:final_ts_gh200}
\end{figure}

Figure~\ref{fig:final_ts_gh200} shows the performance on the GH200 system. cuSOLVER is not shown because it does not support this range. The curve of KBLAS stops at 1,000 for double precision because the library only supports this data type if the number of rows is at most 1,024. For all data types, using the QR factorization yields a performance advantage which diminishes as the number of rows increases. Note that when the QR factorization is used, MAGMA always computes the SVD of a small $n\times n$ matrix, where $n = 16$ in this experiment, so the timing of the SVD stage of the solver depends only on the data type, not on the number of rows in the matrix. Therefore, the time-to-solution grows only because of the increased time of the QR factorization and the postprocessing stage of applying $Q$ to recover the left singular vectors. 

As mentioned, the MAGMA SVD solver uses the existing batch QR factorization in MAGMA~\cite{abdelfattah2022batch}, which implements the standard Householder QR factorization of the LAPACK \verb+GEQRF+ routine. If the matrix has few rows, the MAGMA batch QR implementation applies optimizations that maximize data reuse by caching relatively large parts of each matrix in the fast memory levels of the GPU. This is no longer possible when the number of rows increases, and the factorization automatically switches to a more general implementation. This explains the steps in the execution time of the MAGMA QR-based solver---for example at sizes 1,000 and 1,700 for double data type. More efficient algorithms exist for computing the QR factorizations of tall-skinny matrices, but they are not available in MAGMA and their implementation is beyond the scope of this work.

We observe that MAGMA is uniformly faster than the CPU implementation for all data types, achieving speedups between $2.6\times$ and $10.7\times$. KBLAS has a performance advantage in single precision for relatively small sizes, but it does not support the computation of the right singular vectors. In double precision, the MAGMA SVD solver is 1.2--1.5$\times$ faster than KBLAS. 

\begin{figure}[!htb]
\centering
\includegraphics[width=0.49\linewidth]{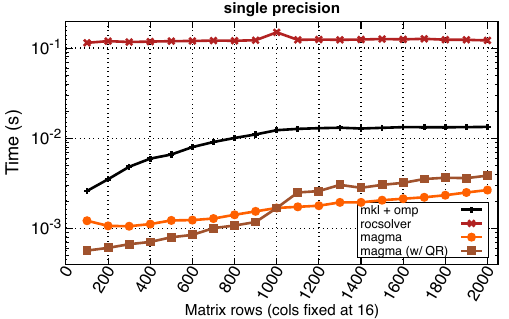}
\includegraphics[width=0.49\linewidth]{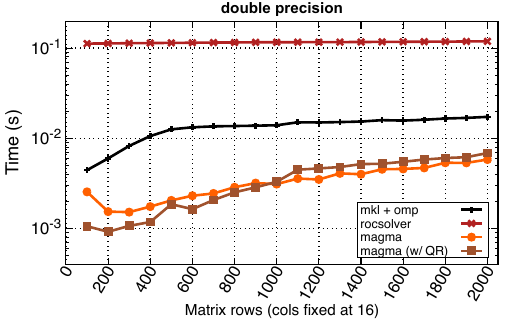}\\
\vspace{1em}
\includegraphics[width=0.49\linewidth]{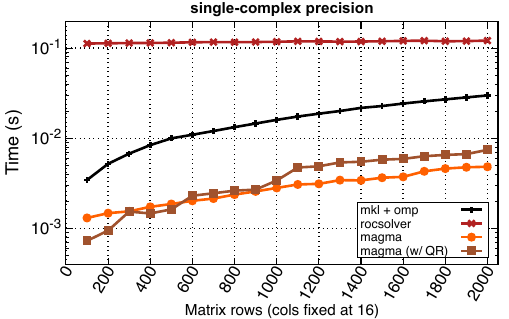}
\includegraphics[width=0.49\linewidth]{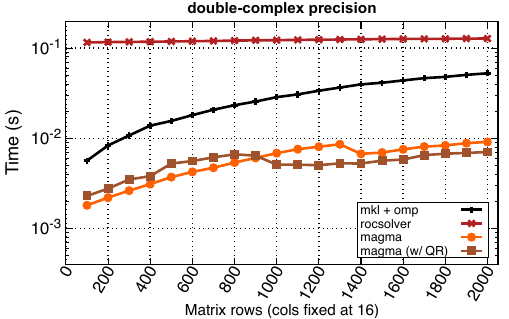}
\caption{Performance of the batch SVD on the MI300A APU. Results are shown for batches of 1,000 tall-skinny matrices.}
\label{fig:final_ts_mi300a}
\end{figure}

On the MI300A system, we observe a mixed behavior for the MAGMA QR-based solver, largely because of the batch QR factorization. The relatively small shared memory on the GPU limits the occupancy of the overall solver, and forces the QR factorization to switch to the generic implementation earlier than on the GH200 system. This leads to an asymptotic slowdown compared with the direct SVD solver, except for the double-complex data type, which shows a different behavior. In this case, the greater storage requirements increases the pressure on the shared memory for both MAGMA approaches. However, the asymptotic behavior shows a slight advantage for the QR-based solver. 
Considering the best timings out of the two MAGMA approaches, we observe speedups between $3\times$ and $9.4\times$ across all precisions. Compared with rocSOLVER, the MAGMA solvers are one to two orders of magnitude faster, achieving speedups in the range of 
45.7--$202.5\times$ for single precision,  
20.5--$123.1\times$ for double precision, 
25.1--$155.1\times$ for single-complex precision, and 
18.1--$63.8\times$ for double-complex precision.
  
\section{Conclusion and Future Work}
\label{sec:conclusion}

We presented a high‑performance batch SVD solver targeting modern GPU architectures. The proposed solver supports all four standard LAPACK data types and can compute singular values and both left and right singular vectors. The implementation is based on the one‑sided Jacobi algorithm and is designed to exploit fine‑grained parallelism across large batches of relatively small matrices. Starting from a blocked baseline design, we introduced a sequence of algorithmic and implementation optimizations that improved the time‑to‑solution without compromising numerical accuracy.
Numerical results on both NVIDIA and AMD systems demonstrate that the resulting solver is numerically robust and outperforms existing GPU alternatives across a wide range of matrix sizes, shapes, and numerical properties. The proposed design achieves significant speedups over vendor libraries and open‑source batch SVD solvers.

Beyond producing a batch SVD solver with competitive performance, our work highlights several design principles that are broadly applicable to batch linear algebra on accelerators. We show that Jacobi-type methods can be an effective option for small and medium problem sizes, we demonstrate the importance of minimizing data movement through kernel fusion and register reuse, and the potential of masked batch execution to address variability of convergence speed within a batch. Many of these ideas extend naturally to other iterative factorization algorithms.

\begin{acks}
The early stages of this work were supported by the Exascale Computing Project (17-SC-20-SC), a collaborative effort of the U.S. Department of Energy Office of Science and the National Nuclear Security Administration. 
This material is based upon work partially supported by the U.S. Department of Energy, Office of Science, Office of Advanced Scientific Computing Research, Next-Generation Scientific Software Technologies program, under contract number DE-AC02-06CH11357.
The second author is a member of the Istituto Nazionale di Alta Matematica.
We thank the Performance Research Laboratory at the University of Oregon for using resources on the Frank cluster. We also thank the Experimental Computing Laboratory at Oak Ridge National Laboratory and Livermore Computing at Lawrence Livermore National Laboratory for access to compute resources. 
We would like to thank Vjeran Hari for extensive feedback on an earlier version of this work.
Finally, this work is dedicated to Stan Tomov (1971--2024), the founder of the MAGMA library, and a prominent figure in high-performance computing.  
\end{acks}

\bibliographystyle{ACM-Reference-Format}
\bibliography{bib/batched, bib/magma, bib/mxp, bib/ref, bib/others, bib/svd, bib/references}

\end{document}